\documentclass[superscriptaddress,amsmath,amssymb, aps, prx,longbibliography,twocolumn, footinbib]{revtex4-2}

\usepackage{upgreek}
\usepackage{color}
\usepackage{graphicx}
\usepackage{nccmath}
\usepackage{bm}
\usepackage{xr-hyper}
\usepackage{xr}
\usepackage{hyperref}
\hypersetup{colorlinks,breaklinks,
            urlcolor=[rgb]{0,0,0.64},
            linkcolor=[rgb]{0,0,0.64},
            citecolor=[rgb]{0,0,0.64},
            filecolor=[rgb]{0,0,0.64}}
\usepackage[mathlines]{lineno}
\usepackage{dcolumn}

\makeatletter
\newcommand*{\addFileDependency}[1]{
  \typeout{(#1)}
  \@addtofilelist{#1}
  \IfFileExists{#1}{}{\typeout{No file #1.}}
}
\makeatother
 
\newcommand*{\myexternaldocument}[1]{%
    \externaldocument{#1}%
    \addFileDependency{#1.tex}%
    \addFileDependency{#1.aux}%
}

\usepackage{mystyle}

\myexternaldocument{supplement}

\newcommand{\Harvard}{Department of Physics, Harvard University, Cambridge, Massachusetts 02138, USA.}

\newcommand{\ETH}{Institute for Theoretical Physics, ETH Zurich, 8093 Zurich, Switzerland.}

\newcommand{\PurduePhys}{Department of Physics and Astronomy, Purdue University, West Lafayette, Indiana 47907, USA.}

\newcommand{\PurdueChem}{Department of Chemistry, Purdue University, West Lafayette, Indiana 47907, USA.}

\newcommand{\HarvardEng}{John A. Paulson School of Engineering and Applied Sciences, Harvard University; Cambridge,
MA 02138, USA.}

\newcommand{\HarvardChem}{Department of Chemistry and Chemical Biology, Harvard University; Cambridge, MA 02138,
USA.}

\newcommand{\JapanFirst}{Research Center for Materials Nanoarchitectonics, National Institute for Materials Science,  1-1 Namiki, Tsukuba 305-0044, Japan.}

\newcommand{\JapanSecond}{Research Center for Electronic and Optical Materials, National Institute for Materials Science, 1-1 Namiki, Tsukuba 305-0044, Japan.}

\begin{document}

\title{Optical signatures of interlayer electron coherence
in a bilayer semiconductor}

\author{Xiaoling~Liu}
\thanks{These authors contributed equally to this work.}
\affiliation{\Harvard}
\author{Nadine~Leisgang}
\thanks{These authors contributed equally to this work.}
\affiliation{\Harvard}
\author{Pavel~E.~Dolgirev}
\thanks{These authors contributed equally to this work.}
\affiliation{\Harvard}

\author{Alexander~A.~Zibrov}
\affiliation{\HarvardChem}
\affiliation{\Harvard}

\author{Jiho~Sung}
\affiliation{\HarvardChem}
\affiliation{\Harvard}

\author{Jue~Wang}
\affiliation{\HarvardChem}
\affiliation{\Harvard}

\author{Takashi~Taniguchi}
\affiliation{\JapanFirst}

\author{Kenji~Watanabe}
\affiliation{\JapanSecond}

\author{Valentin~Walther}
\affiliation{\PurduePhys}
\affiliation{\PurdueChem}
\affiliation{\Harvard}

\author{Hongkun~Park}
\affiliation{\HarvardChem}
\affiliation{\Harvard}

\author{Eugene~Demler}
\affiliation{\ETH}

\author{Philip~Kim}
\affiliation{\Harvard}
\affiliation{\HarvardEng}

\author{Mikhail~D.~Lukin}\email[Correspondence to: ]{lukin@physics.harvard.edu}
\affiliation{\Harvard}

\date{\today}

\maketitle

\textbf{
Emergent strongly-correlated electronic phenomena in atomically-thin transition metal dichalcogenides are an exciting  frontier in condensed matter physics, with examples ranging from bilayer superconductivity~\cite{zhao2023evidence} and electronic Wigner crystals~\cite{smolenski2021signatures,zhou2021bilayer} to the ongoing quest for exciton condensation~\cite{wang2019evidence,ma2021strongly,shi2022bilayer}.
Here, we experimentally investigate the properties of indirect excitons in naturally-grown MoS$_2$-homobilayer, integrated in a dual-gate device structure allowing independent control of the electron density and out-of-plane electric field. 
Under conditions when electron tunneling between the layers is negligible~\cite{pisoni2019absence}, upon electron doping the sample, we observe that the two excitons with opposing dipoles hybridize, displaying unusual behavior distinct from both conventional level crossing and anti-crossing. 
We show that these observations can be explained by static random coupling between the excitons, which increases with electron density and decreases with temperature.
We argue that this phenomenon is indicative of a spatially fluctuating order parameter in the form of interlayer electron coherence, a theoretically predicted many-body state~\cite{zheng1997exchange} that has yet to be unambiguously established experimentally outside of the quantum Hall regime~\cite{sarma2008perspectives,spielman2000resonantly,kellogg2004vanishing,kellogg2002observation,spielman2001observation,fertig1989energy,shi2022bilayer}.
Implications of our findings for future experiments and quantum optics applications are discussed. 
}

\begin{figure*}[btp!]
\centering
\includegraphics[width=180mm]{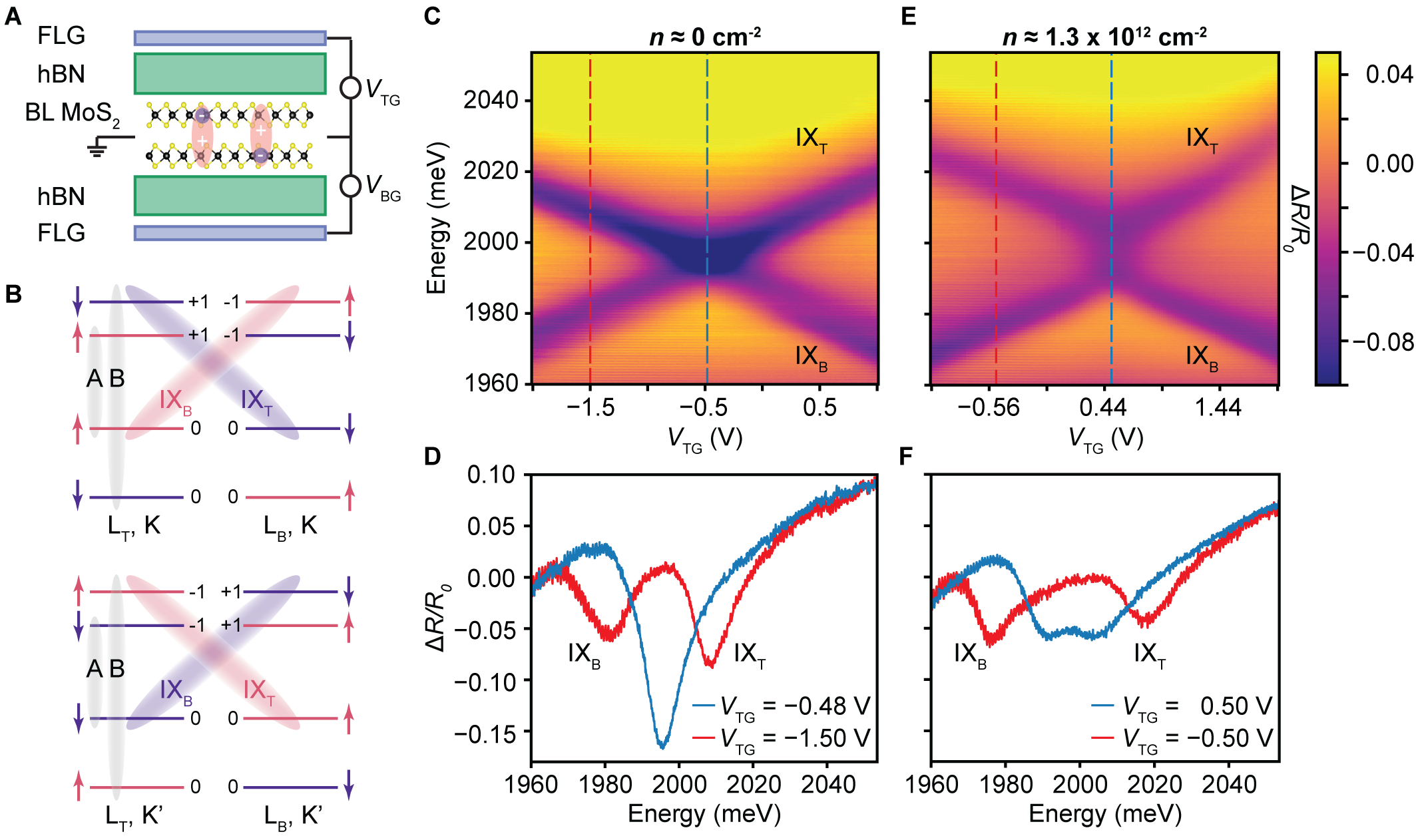}
\caption{DC Stark effect of interlayer excitons (IXs). \textbf{A} Schematic of a 
dual-gated 2H-stacked MoS$_2$-homobilayer encapsulated with hexagonal boron nitride (hBN). Tuning of the top and bottom gates, composed of a few layers of graphene (FLG), allows for independent control of the total electron density $n$ and out-of-plane electric field $E_z$. Interlayer excitons, highly sensitive to $E_z$ due to their large dipole moments, are also depicted.
\textbf{B} Schematic electronic bandstructure near the $K$- (top) and $K'$-valleys (bottom) showing the relevant excitonic levels, electron spin, and corresponding azimuthal quantum numbers (AQNs) of the electronic bands, which determine optical selection rules. 
\textbf{C} In the undoped case $n = 0$, the energies of interlayer excitons shift linearly with $E_z$, as can be seen in the simple crossing of exciton branches in the measured reflectance map $\Delta R/R_0$ at $T = 8\,$K.
\textbf{D} The system exhibits two well-separated branches at a finite $E_z\neq 0$, becoming degenerate at $E_z = 0$, with doubled oscillator strength.  
\textbf{E} DC Stark effect for the doped sample with $n\approx 1.3\times 10^{12}\,$cm$^{-2}$, showing that the simple crossing in \textbf{C} turns into a stochastic avoided crossing (Fig.~\ref{Fig2}). 
\textbf{F} Linecut at $V_{\rm TG} = 0.50\,$V, corresponding to $E_z = 0$, displays a broad feature with reduced relative amplitude compared to the undoped case in \textbf{D}.
 }
\label{Fig1}
\end{figure*}

Transition metal dichalcogenides (TMDs) are direct-gap semiconductors, which can host optically bright excitons corresponding to Coulomb-bound electron-hole pairs.
Due to the two-dimensional (2D) nature of TMDs, along with the large effective masses of electrons and holes and small dielectric permittivity of the surrounding medium, excitons are tightly confined, with the Bohr radius substantially smaller than the typical separation between doped charges~\cite{he2014tightly}.
These features make excitons in TMDs a promising tool for optical probing of many-body electron correlations. 
In particular, understanding the exciton fine structure of a doped sample has proven pivotal for a number of recent discoveries.
Examples range from investigating polaronic dressing effects, which manifest through exciton line splitting into attractive and repulsive branches~\cite{sidler2017fermi}, to probing correlated many-body phases using excited-state spectroscopy~\cite{xu2020correlated}, to observing electron crystalline states via umklapp scattering~\cite{smolenski2021signatures}, and to studying the rich  magnetic properties of TMDs~\cite{roch2019spin,ciorciaro2023kinetic,sung2023observation}.
While most of the prior studies focused on intralayer excitons, where both the exciton electron and hole reside in the same TMD layer, bilayer TMDs can host interlayer excitons (Fig.~\ref{Fig1}A,B), where the electron and hole are separated across the two layers~\cite{jiang2021interlayer}.
However, interlayer excitons typically have weak optical transition dipole moments, posing challenges for optical measurements.
In materials like MoS$_2$-homobilayers, intra- and interlayer excitons strongly hybridize~\cite{gerber2019interlayer,leisgang2020giant,deilmann2018interlayer}, making interlayer excitons optically bright and enabling their use for optical probing of electronic correlations.

Here, we experimentally investigate the properties of indirect excitons in naturally-grown MoS$_2$-homobilayer, integrated into a dual-gate device structure (Fig.~\ref{Fig1}A) whereby the top and bottom gate voltages, $V_{\rm TG}$ and $V_{\rm BG}$, are simultaneously used to independently control the out-of-plane electric field $E_z$ and the electron density $n$ in the sample (see Methods). 
The interlayer excitons (IXs) have large permanent electric dipole moments $\pm d_z$ (Fig.~\ref{Fig1}A), which make them highly sensitive to $E_z$. This can be studied by measuring reflectance contrast spectra $(R- R_0)/R_0 = \Delta R/R_0$ using a weak, incoherent white light source, where $R$ is the reflectance obtained on the bilayer MoS$_2$ flake and $R_0$ is the reference spectrum at a high doping level (see Methods). 
Figure~\ref{Fig1}C shows the undoped case ($n = 0$), illustrating the DC Stark effect, where the two interlayer excitons with opposite dipoles shift linearly with $E_z$ and cross at $E_z = 0$ ($V_{\rm TG} \approx -0.48\,\text{V}$). 
The degeneracy point $E_z = 0$ is characterized by the amplitude doubling in the reflectance contrast spectrum of interlayer excitons, Fig.~\ref{Fig1}D (blue curve). 
Surprisingly, when the sample is doped ($n \approx 1.3\times 10^{12}\,$cm$^{-2}$, as extracted from simulations based on a simple capacitance model in Sec.~I in SI), the simple crossing in Fig.~\ref{Fig1}C turns into an elongated shape shown in Fig.~\ref{Fig1}E.
This effect is highly reproducible across different collection light spots within the same sample, as well as in other similar devices (Sec.~II in SI). 
The putative degeneracy point $E_z = 0$ ($V_{\rm TG} \approx 0.50\,$V), no longer exhibits the amplitude doubling (Fig.~\ref{Fig1}F). Instead, we observe a broadened feature with the overall amplitude roughly the same as that of individual interlayer excitons.

\begin{figure*}[btp]
\centering
\includegraphics[width=180mm]{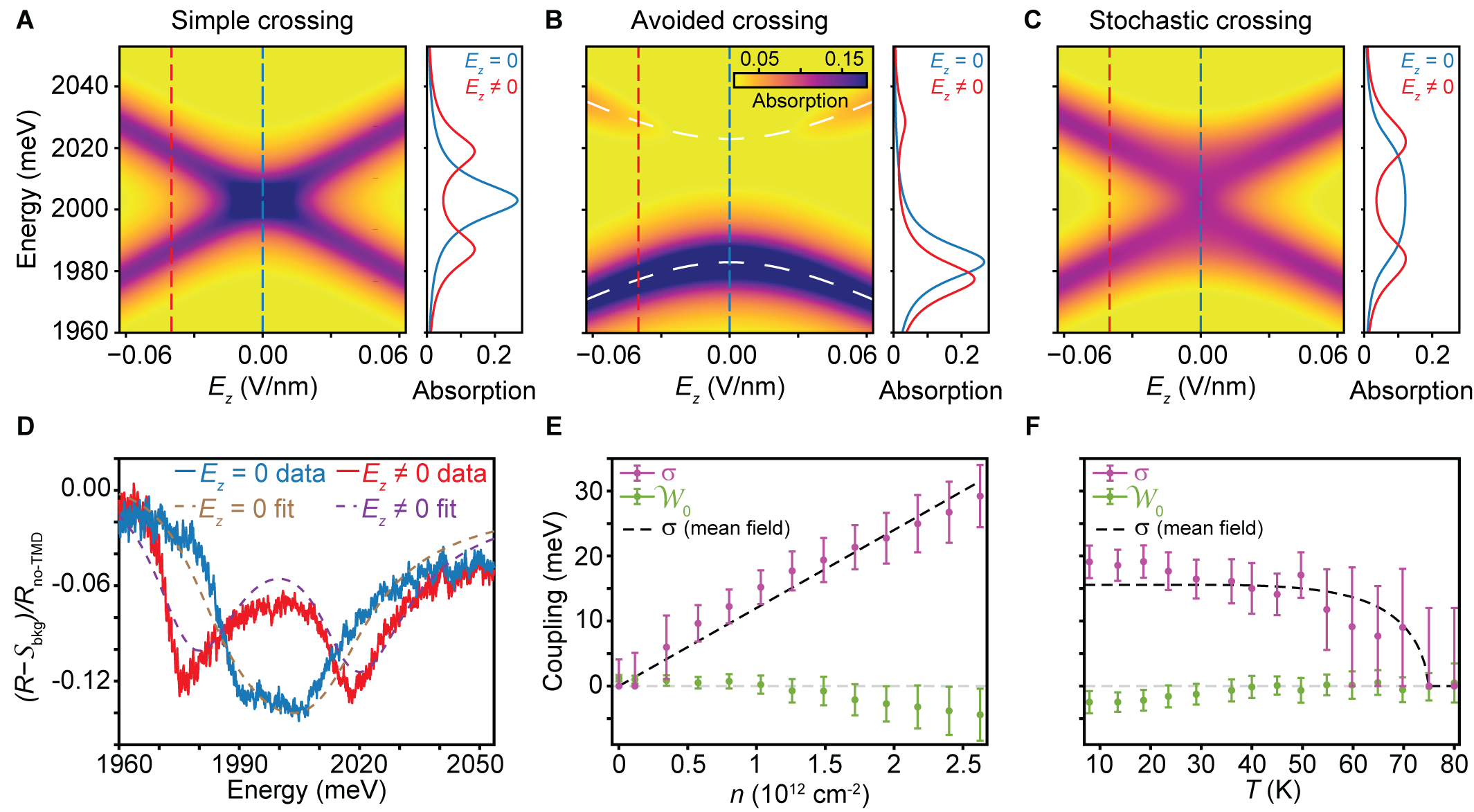}
\caption{Stochastic interlayer exciton hybridization. Simulated absorption map exhibits a simple crossing \textbf{A} as in Fig.~\ref{Fig1}C when the two excitons are uncoupled (${\cal W}_0 = 0,\, \sigma = 0$ in Eq.~\eqref{eqn:homog_main}), an avoided crossing \textbf{B} with asymmetry in the intensities of the two branches when the excitons are hybridized (${\cal W}_0 = -20\,\text{meV},\, \sigma = 0$), and a stochastic crossing \textbf{C} reminiscent of Fig.~\ref{Fig1}E when the exciton coupling has a static, random character (${\cal W}_0 = 0,\, \sigma = 20\,\text{meV}$). 
\textbf{D} The measured reflectance contrast spectra are analyzed using a few parameter-fit based on the model of stochastic coupling in Eq.~\eqref{eqn:homog_main}; shown are two linecuts at $n \approx 1.2\times 10^{12}\,$cm$^{-2}$ corresponding to zero (blue curve) and nonzero electric fields (red curve), respectively. Such a fit (dashed lines) quantitatively captures both the linear Stark effect as well as the stochasticity of the interlayer exciton hybridization.
Here, $R_{\rm bkg}$ is the fitted reflectance encoding background effects, while $R_{\rm no-TMD}$ is the measured reflectance at an optical spot away from the bilayer (Sec.~VIII in SI). 
\textbf{E,F} Evolution of ${\cal W}_0$ and ${\sigma}$ with the electron density $n$ at $T$ $\approx$ 8~K (\textbf{E}) and temperature $T$ at $n \approx 1.3\times 10^{12}\,$cm$^{-2}$ (\textbf{F}). We find that both $|{\cal W}_0|$ and ${\sigma}$ increase (decrease) with increasing $n$ ($T$), indicating a stronger hybridization between excitons at higher electron densities and lower temperatures. Dashed lines in \textbf{E,F} represent mean-field trends for the stochastic variance $\sigma$ (Sec.~IX in SI).
}
\label{Fig2}
\end{figure*}

To understand these observations, we consider a simple model of two coupled harmonic oscillators describing the excitonic polarization response to the probe electric field ${\cal E}(t)$:
\begin{gather}
     i\hbar \partial_t X_{\rm T} = \omega_{\rm T} X_{\rm T} - i \gamma_{\rm T} X_{\rm T} + {\cal W} X_{\rm B}  - d_{\rm T} {\cal E}(t), \label{eqn:osc}
\end{gather}
and a similar equation holds for ${X}_{\rm B}$. 
Here, the variable ${X}_{\rm T/B}$ represents the polarization oscillations associated with the interlayer exciton  $\rm IX_{T/B}$  (Fig.~\ref{Fig1}B), with the subscript referring to the layer of the electron; $\omega_{\rm T/B} = \pm d_zE_z$ is the energy relative to the degeneracy point $E_z = 0$; $\gamma_{\rm T/B}$ is the total respective exciton decay rate; $d_{\rm T/B}$ refers to the corresponding transition dipole moment; and ${\cal W}$ is the coupling strength between the two interlayer excitons, which we introduced for reasons that will become clear shortly. 
Figure~\ref{Fig2}A depicts a simulated absorption map ${\rm Im}[\chi(\omega)]$, where $\chi(\omega)$ is the polarization response function of the sample (Sec.~VI in SI). 
This simulation corresponds to a simple crossing with ${\cal W} = 0$ and closely resembles the measured signal for $n= 0$ in Fig.~\ref{Fig1}C. 
For ${\cal W} \neq 0$, an avoided crossing occurs, characterized by an asymmetry in intensities between the upper and lower exciton branches (Fig.~\ref{Fig2}B) -- this effect is attributed to constructive/destructive interference in the photon emission process of the corresponding exciton branches (Sec.~VII in SI).

While we observe a slight asymmetry in intensities in Fig.~\ref{Fig1}E, the overall elongated shape at high doping is clearly not captured by either conventional level crossing (Fig.~\ref{Fig2}A) or anti-crossing (Fig.~\ref{Fig2}B). 
Instead, we find that the experimental data are well represented by a model that incorporates ensemble averaging over the coupling ${\cal W}$, treated as a random, static variable distributed as:
\begin{gather}
    \langle {\cal W}\rangle = {\cal W}_0,\qquad  \delta {\cal W} = ({\cal W} - {\cal W}_0) \in [-\sigma, \sigma]. 
        \label{eqn:homog_main}
\end{gather}
Here, ${\cal W}_0$ is the mean coupling strength, while $\sigma$ encodes the variance. The corresponding simulated absorption map (Sec.~VI in SI), shown in Fig.~\ref{Fig2}C, qualitatively agrees with Fig.~\ref{Fig1}E, 
capturing two distinctive features: (i) a near equal intensity distribution between the upper and lower interlayer exciton branches and (ii) a plateau-like flattening of the signal along $E_z = 0$.
For this reason, the elongated shape in Fig.~\ref{Fig1}E is further referred to as stochastic anti-crossing. 
We emphasize the importance of the static character of the random coupling ${\cal W}$. If ${\cal W}$ were instead a time-dependent Markovian variable, its effects would be fully accounted for through a modification of the decay rates $\gamma_{\rm T}$ and $\gamma_{\rm B}$ (Sec.~VI in SI).

Using the model in Eqs.~\eqref{eqn:osc} and~\eqref{eqn:homog_main},
we analyze the experimental data obtained under a variety of different conditions including different temperatures and dopings. 
Specifically, we simultaneously fit the full reflectance maps (Fig.~\ref{Fig1}E) with a few-parameter model (Sec.~VIII in SI), which incorporates substrate reflectance effects and characterizes the interlayer excitons via six parameters: ${\cal W}_0$, $\sigma$, $\gamma = \gamma_{\rm T} = \gamma_{\rm B}$, $d = d_{\rm T} = d_{\rm B}$, $d_z$, and $\omega_0$, which is the bare interlayer exciton energy at $E_z = 0$. 
The density and temperature behavior obtained from this analysis, shown in Fig.~\ref{Fig2}E,F, reveals that the static stochastic variance $\sigma$ increases with increasing $n$ and decreases with increasing $T$. 
The data also point at the development of a nonzero mean coupling ${\cal W}_0\neq 0$ (Sec.~VII in SI), which is consistently found to be relatively small $|{\cal W}_0|\ll \sigma$. The mean coupling ${\cal W}_0$ roughly follows the trend of $\sigma$, but for $n \approx 1.3\times 10^{12}\,$cm$^{-2}$ vanishes at around $T \approx 40\,$K, while $\sigma$ persists up to about $T \approx 75\,$K (Sec.~IV in SI).

\begin{figure}[btp!]
\centering
\includegraphics[width=85mm]{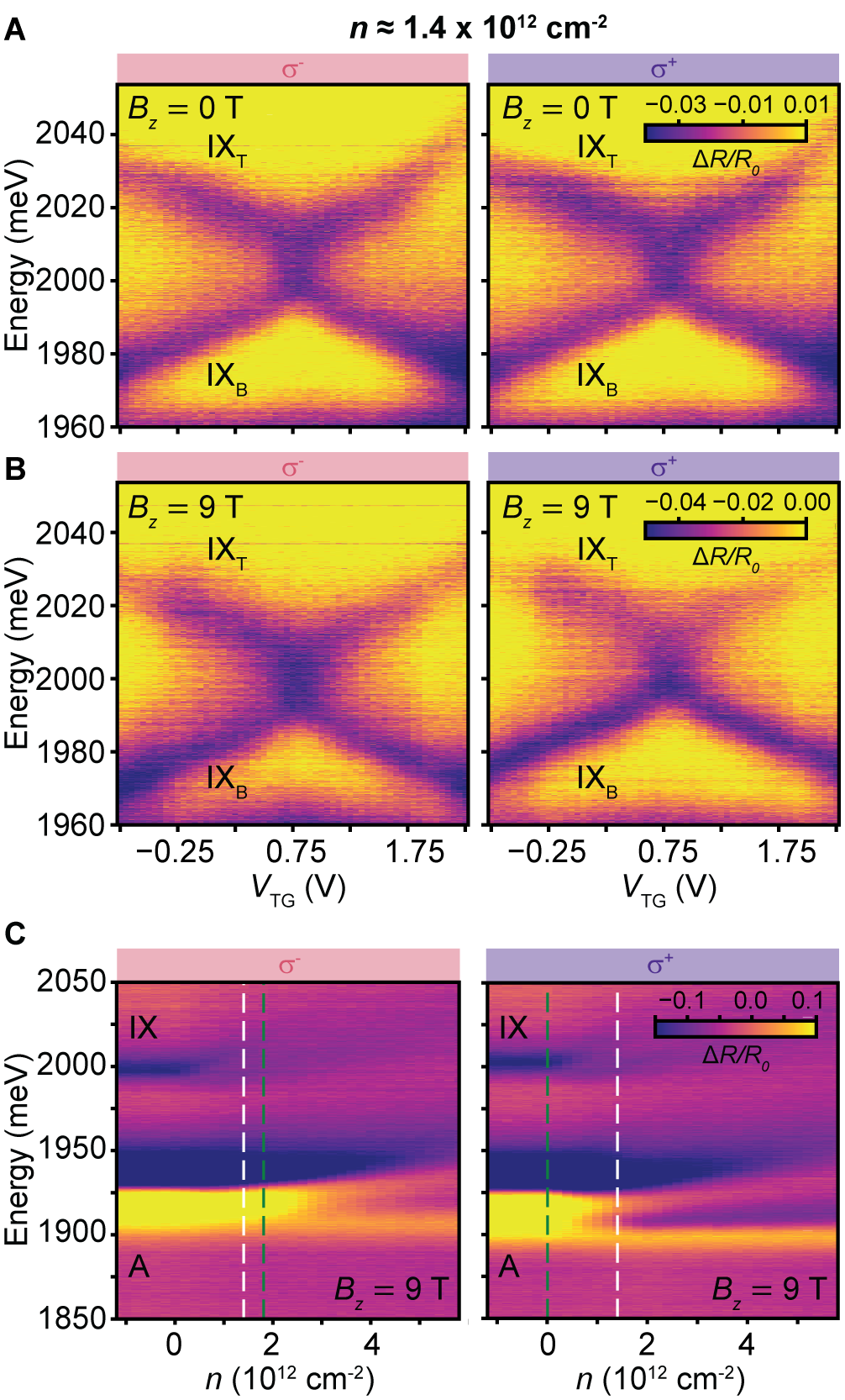}
\caption{ Magnetic field and polarization-resolved properties. 
\textbf{A,B} Electric-field sweeps at $n\approx 1.4\times 10^{12}\,$cm$^{-2}$ illustrate the similar appearance of the stochastic anti-crossing for both light polarizations and for both $B_z = 0\,$T \textbf{A} and $B_z = 9\,$T \textbf{B}.
At $B_z = 9\,$T, the $\sigma^+$-measurements reveal a small intensity asymmetry between the lower and upper exciton branches, suggesting a slight development of ${\cal W}_0$ for this light polarization.
\textbf{C} At $B_z = 9\,$T and $n\approx 1.4\times 10^{12}\,$cm$^{-2}$ (white dashed lines), conduction band electrons are expected to be fully spin-polarized. This is supported by density sweeps at $E_z = 0$ of the $A$-exciton, where the onset of the attractive polaron branch for $\sigma^-$-polarized light (primarily probing spin-$\uparrow$ electrons) is delayed compared to the $\sigma^+$-polarized one (essentially sensitive to spin-$\downarrow$ electrons) -- these onsets are indicated with green dashed lines.  
}
\label{Fig3}
\end{figure}

To gain further insights into the nature of this exciton hybridization, we examine both the valley and spin structure of indirect excitons, illustrated in Fig.~\ref{Fig1}B.
With two inequivalent valleys, associated with the $K$- and $K'$-points of the hexagonal Brillouin zone (BZ), there are four relevant, optically bright interlayer excitons in total: two excitons with opposite out-of-plane dipoles per valley.
The 2$H$-stacked MoS$_2$-homobilayer exhibits ${\cal C}_3$-rotational symmetry, assigning azimuthal quantum numbers (AQNs) to each of its electronic bands (Fig.~\ref{Fig1}B). The AQNs of the valence bands are zero, allowing holes to tunnel between layers. 
Conversely, the AQNs of the conduction bands in the same valley are opposite, which is the fundamental reason that prevents electron tunneling~\cite{gong2013magnetoelectric,pisoni2019absence,gerber2019interlayer} and, thus, naively should prevent interlayer exciton hybridization.
The AQNs also dictate the optical selection rules for excitons~\cite{cao2012valley,xiao2012coupled,gerber2019interlayer}: an electron with AQN $+1$ ($-1$) corresponds to an exciton coupling to $\sigma^+$- ($\sigma^-$-) polarized light.

One notable feature of MoS$_2$-homobilayers is their small conduction-band spin-orbit splitting of a few meV, which could result in spin polarization, though not necessarily valley polarization, of conduction-band electrons via an out-of-plane magnetic field $B_z$.
This expectation is corroborated by our measurements of polarization-resolved reflection contrast spectra of the intralayer $A$-exciton at $B_z = 9\,$T, $E_z = 0$, and varying $n$ (Fig.~\ref{Fig3}C).
We observe that the attractive polaron (AP) branch for $\sigma^-$-polarized light, predominantly sensing spin-$\uparrow$ electrons, emerges at a higher doping level compared to the $\sigma^+$-polarized one, which primarily probes spin-$\downarrow$ electrons~\cite{roch2019spin}. 
As a result, for electron densities in the asymmetry region between the two AP onsets (green dashed lines in Fig.~\ref{Fig3}C), conduction band electrons become fully spin-polarized. 
Additionally, previous magnetism studies on monolayer MoS$_2$~\cite{roch2019spin} suggest that these spin-polarized electrons remain valley-depolarized.
For one such representative density $n\approx 1.4\times 10^{12}\,$cm$^{-2}$ (white dashed lines in Fig.~\ref{Fig3}C), we find that the stochastic anti-crossing is robustly present for both light polarizations and for both $B_z = 0$ (Fig.~\ref{Fig3}A) and $B_z = 9\,$T (Fig.~\ref{Fig3}B). 
Within the error margin of our analysis (Sec.~VIII in SI), the stochastic variance $\sigma$ is found to be around $15\,$meV across all four panels in Fig.~\ref{Fig3}A,B, while the mean coupling ${\cal W}_0$ is nearly zero throughout. 
A slight departure from this trend is that ${\cal W}_0$  develops by at most $-2\,$meV for $\sigma^+$-polarized light at $B_z = 9\,$T, as evidenced by a small intensity asymmetry between the upper and lower exciton branches in the right panel of Fig.~\ref{Fig3}B. 
The persistent presence of a large $\sigma$ nearly independent of $B_z$ indicates that  
the interlayer exciton hybridization is predominantly agnostic to the electron spin.

We now turn to the theoretical interpretation of our observations.  
The stochastic anti-crossing in Fig.~\ref{Fig1}E can be attributed to intravalley and/or intervalley interlayer exciton hybridization. 
The modest asymmetry in the lower and upper branches (associated with 
a small nonzero mean coupling ${\cal W}_0 \neq 0$ in the model given by Eqs.~\eqref{eqn:osc} and~\eqref{eqn:homog_main}) is likely due to  the intervalley scenario, as interlayer excitons within any of the two valleys have opposite AQNs and the optical interference effects that give rise to ${\cal W}_0 \neq 0$ are suppressed for excitons with opposite polarizations (Sec.~VII in SI).
In contrast, the stochastic variance $\sigma \neq 0$ is compatible with both scenarios (Sec.~VII in SI), suggesting that both types of hybridization can play a role.

Hybridization between intervalley interlayer excitons with opposing dipoles is allowed from a symmetry perspective, as these excitons, such as ${\rm IX}_{\rm T,K'}$ and ${\rm IX}_{\rm B,K}$ (depicted in red in Fig.~\ref{Fig1}B), have the same AQNs.
Even without electron doping the sample, these could hybridize with each other via direct Coulomb interactions: either via exciton exchange~\cite{pikus1971exchange,yu2014dirac}, expected to be weak because of the reduced transition dipole moment of interlayer excitons compared to intralayer ones, or a process involving the scattering of both the ${\rm IX}_{\rm T,K'}$-exciton electron and hole across the TMD BZ, which is suppressed because it occurs at a large momentum $\bm K-\bm K'$ and involves electron and hole layer switching (Sec.~X in SI). 
Thus, such direct coupling is expected to be weak, consistent with $|{\cal W}_0|\lesssim 2\,$meV for $n = 0$ (Fig.~\ref{Fig2}E).
Doping the sample could enhance such hybridization mechanisms via simple effects such as polaronic dressing or Fermi sea fluctuations, possibly explaining the emergence of nonzero mean coupling ${\cal W}_0\neq 0$ and the density trend in Fig.~\ref{Fig2}E (such dynamical electron-enhanced exciton hybridization is still expected to be suppressed, consistent with our measurements in Fig.~\ref{Fig2}E,F, as further discussed in Sec.~X in SI).
The intensity asymmetry in the right panel of Fig.~\ref{Fig3}B could arise from the presence of doped electrons indistinguishable from the corresponding exciton electron.
Increasing temperature weakens polaronic dressing effects~\cite{mulkerin2023exact} and increases exciton scattering off phonons~\cite{selig2016excitonic}, which reduces exciton wave-function overlaps. 
Both effects may contribute to explaining decreasing $|{\cal W}_0|$ with increasing $T$ as observed in Fig.~\ref{Fig2}F.

\begin{figure*}[btp!]
\centering
\includegraphics[width=180mm]{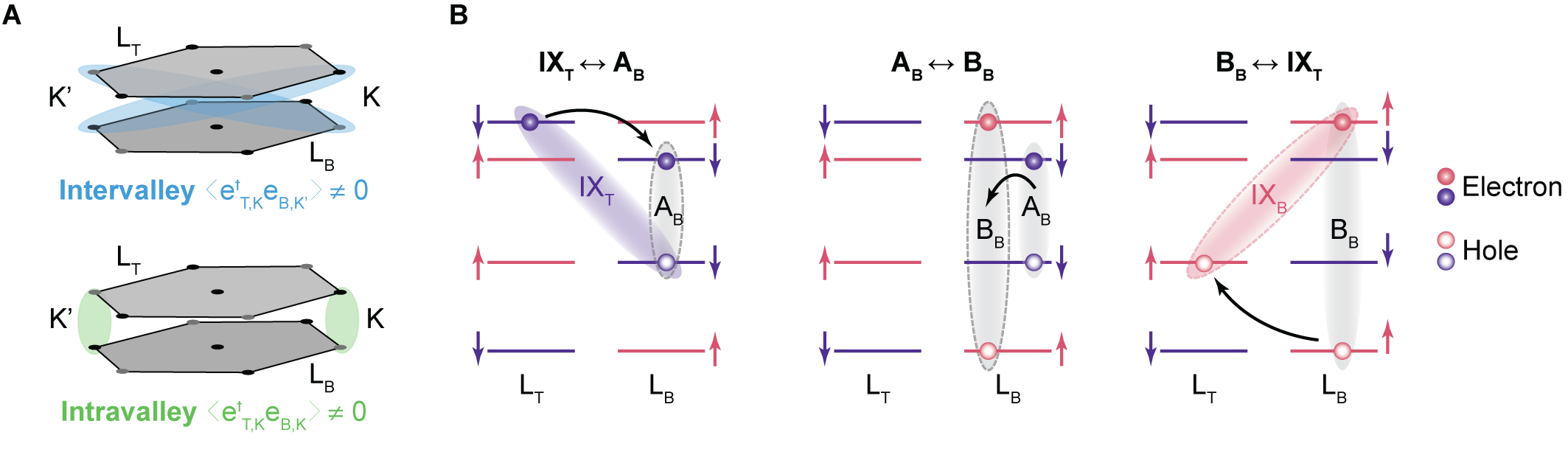}
\caption{  
Coulomb-mediated mechanism of interlayer exciton hybridization.
\textbf{A} The electronic many-body state can exhibit interlayer electron coherence with intervalley (top) or intravalley correlations (bottom). 
\textbf{B} Intravalley coherence leads to an effective order-parameter-induced electron tunneling-like process (left), resulting in hybridization between ${\rm IX}_{\rm T}$- and $A_{\rm B}$-excitons (shown is the $K$-valley). The $A_{\rm B}$-exciton couples to the ${\rm IX}_{\rm B}$-state via exciton exchange (middle) followed by hole tunneling (right), thereby hybridizing the two interlayer excitons.
}
\label{Fig4}
\end{figure*}

At the same time, the emergence of the stochastic variance $\sigma$ involves quasi-static processes, which are beyond the simple dynamical processes mentioned previously, especially given the large values of $\sigma$ in Fig.~\ref{Fig2}E,F.
Moreover, the effects of quenched disorder or charge traps should be mitigated via electron screening, particularly because strongly-interacting regimes in TMDs can be achieved at significantly higher electron densities than in conventional semiconductors~\cite{smolenski2021signatures,zhou2021bilayer,sung2023observation}.
Experimentally, we observed $\sigma$ increases as $n$ increases, which invalidates disorder-induced scenarios.
Instead, $\sigma$ could originate from a correlated many-body state that develops an order parameter $\Delta$, in which case the observed stochastic behavior is attributed to quasi-static spatial fluctuations of this order parameter (Fig.~\ref{Fig4}C).
In particular, one potential candidate is interlayer electron coherence, corresponding to an exchange instability akin to the typical emergence of ferromagnetism. 
This correlated state has been proposed theoretically~\cite{zheng1997exchange} and experimentally established in quantum Hall bilayers~\cite{sarma2008perspectives,spielman2000resonantly,kellogg2004vanishing,kellogg2002observation,spielman2001observation,fertig1989energy,shi2022bilayer,fertig2005coherence}, where the strong magnetic field quenches the electron kinetic energy and, thus, favors an ordered phase, but it has not yet been conclusively observed at $B_z = 0$.
Such a state requires (i) strong Coulomb interactions $1 \ll r_s \equiv m^* e^2/(4 \pi \varepsilon_0 \varepsilon \hbar^2 \sqrt{\pi n})$ ($m^*$ and $\varepsilon$ are the effective electron mass and permittivity of the surrounding medium, respectively), (ii) the absence of electron tunneling, and (iii) a small interlayer separation $lk_F \ll 1$ ($k_F$ is the Fermi momentum and $l\simeq 0.6\,$nm is the interlayer separation).

Our experimental conditions in the studied MoS$_2$-homobilayer naturally fulfill these stringent prerequisites.
First, the large effective mass $m^* \approx 0.7 m_e$ and the small permittivity of hBN, $\varepsilon \approx 3.76$~\cite{laturia2018dielectric}, result in $r_s \simeq 20$ for $n = 1\times 10^{12}\,$cm$^{-2}$ and $r_s \simeq 11.5$ for $n = 3\times 10^{12}\,$cm$^{-2}$. 
Second, due to the intravalley conduction-band AQNs mismatch in Fig.~\ref{Fig1}B and as experimentally confirmed in Ref.~\cite{pisoni2019absence}, electron tunneling between the layers is intrinsically absent.
Third, we estimate $lk_F \simeq 0.2$ for $n = 3\times 10^{12}\,$cm$^{-2}$. Finally, by studying samples with varying hBN thickness to modulate the strength of Coulomb interactions, we confirm the Coulomb origin of the studied phenomenon (Sec.~III in SI).

The putative emergence of interlayer electron coherence may manifest as the stochastic anti-crossing via a Coulomb-mediated mechanism in Fig.~\ref{Fig4}B, consistent with and potentially explaining our observations. 
In conventional semiconductor double quantum wells, the order parameter is associated with the spontaneous breaking of U(1) symmetry, corresponding to in-plane rotations of the layer pseudospin -- the up and down directions of the pseudospin represent the top and bottom layers, respectively (for simplicity, we omit discussion of electron spin).
In MoS$_2$-homobilayers, the presence of two valleys enriches this symmetry to U(1)$\times$SU(2), where the SU(2) part is related to valley pseudospin rotations (Sec.~X in SI discusses the approximate nature of this U(1)$\times$SU(2) symmetry in TMDs).
This enlarged symmetry places intervalley (Fig.~\ref{Fig4}A, top) and intravalley (Fig.~\ref{Fig4}A, bottom) correlations on equal footing (Sec.~IX in SI).
The significance of intravalley correlations, such as $|\Delta| e^{i\varphi} \sim \langle e^\dagger_{\rm T}e_{\rm B}\rangle \neq 0$ (Fig.~\ref{Fig4}A, bottom), where $e_{\rm B}$ and $e^\dagger_{\rm T}$ are the $K$-valley electron annihilation and creation operators, respectively, is that they lead to strong Coulomb-mediated electron tunneling-like processes~\cite{spielman2000resonantly,lin2022emergence,wen1993tunneling,stern2000dissipationless,stern2001theory,fogler2001josephson} (Sec.~IX in SI).
In momentum space, these can be expressed as (we write only the processes in the $K$-valley):
\begin{align}
    -\sum_{\bm k} t_{\bm k} 
     \hat{e}^\dagger_{\rm B,K}(\bm k)\hat{e}_{\rm T,K}(\bm k) + \text{h.c.},
\end{align}
where the coupling constant $t_{\bm k}$, which provides an effective tunneling-like rate, is determined by both the order parameter amplitude $|\Delta(\bm r)|$ and phase $\varphi(\bm r)$.  
Assuming perfect Hartree-Fock correlations~\cite{zheng1997exchange} and taking into account the angstrom-scale interlayer separation between the TMD layers, we estimate $t_{\bm k}\simeq 100\,$meV for $n = 2\times 10^{12}\,$cm$^{-2}$ (Sec.~IX in SI).
While this estimate is crude, it underscores the significance of the proposed processes.

This electron tunneling-like process gives rise to a hybridization of, for example, ${\rm IX}_{\rm T}$- and ${A}_{\rm B}$-excitons (Fig.~\ref{Fig4}B, left), with the corresponding coupling estimated to be of the order $t_{\rm IX_{\rm T} \leftrightarrow A_{\rm B}}\simeq 85\,$meV for $n = 2\times 10^{12}\,$cm$^{-2}$ (Sec.~IX in SI). 
The ${A}_{\rm B}$-exciton is, in turn, coupled to the ${\rm IX}_{\rm B}$-state via a two-step process shown in the middle and right panels of Fig.~\ref{Fig4}B, involving exciton exchange~\cite{guo2019exchange} (middle panel) followed by hole tunneling~\cite{leisgang2020giant,gerber2019interlayer} (right panel). 
This ${A}_{\rm B}$-${\rm IX}_{\rm B}$ coupling is already established experimentally~\cite{sponfeldner2022capacitively}, and its strength is estimated to be about $t_{A_{\rm B} \leftrightarrow \rm IX_{\rm B}}\simeq 4\,$meV.
Combined, the processes in Fig.~\ref{Fig4}B result in the hybridization of the two interlayer excitons ${\rm IX}_{\rm T}$ and ${\rm IX}_{\rm B}$, with the coupling strength being of the order of $5\,$meV for $n = 2\times 10^{12}\,$cm$^{-2}$ (Sec.~IX in SI). 
Although the above analysis relies on two simplifying assumptions -- perfect Hartree-Fock correlations and a perturbative approach to relating the electronic order parameter to interlayer exciton hybridization -- the estimated number is comparable to the measured values in 
Fig.~\ref{Fig2}E.
Finally, the exciton exchange step in the middle panel of Fig.~\ref{Fig4}B involves flipping both exciton electron and hole spins, indicating that the proposed mechanism is relevant even when conduction-band electrons are spin-polarized by a magnetic field (Fig.~\ref{Fig3}), provided the system remains valley-depolarized~\cite{roch2019spin}.

The corresponding interlayer exciton hybridization $\delta{\cal W}(\bm r)$ is determined by the interlayer electron coherence $|\Delta(\bm r)| e^{i\varphi(\bm r)}$ and thus inherits its spatial inhomogeneities arising from statistical fluctuations of the order parameter phase $\varphi(\bm r)$.
Typically, these fluctuations take the form of vortices; however, in TMDs with the enlarged U(1)$\times$SU(2) symmetry, other meron-like topological defects might be essential~\cite{girvin2019modern}.    
In our experiment, the coupling $\delta{\cal W}(\bm r)$ is spatially averaged over the optical spot size of about $0.5\,\mu$m. 
This size is expected to be much larger than the phase coherence length (at low temperatures, it is on the order of the correlation length of the disorder potential~\cite{rossi2005interlayer}, which we expect to be at most a few hundred nm).  
As a result, upon spatial averaging, the order parameter induced contribution to the
interlayer exciton hybridization vanishes $\langle\delta {\cal W}(\bm r)\rangle = 0$ (see also Sec.~V in SI, where we experimentally explore optical size effects and argue against the phase coherence as the origin of the mean coupling ${\cal W}_0$ in Eq.~\eqref{eqn:homog_main}). 
Nevertheless, an appreciable stochastic variance $\sigma$ in Eq.~\eqref{eqn:homog_main} can develop since it is essentially determined by the order parameter amplitude $|\Delta(\bm r)|$. 
The observed behavior in Fig.~\ref{Fig2}E,F for $\sigma$ is consistent with the development of the amplitude $|\Delta|$ as electron density $n$ increases within the range accessible in our experiment, and its gradual suppression with increasing temperature $T$ until eventual melting -- both these trends are well-captured by the mean-field analysis, as indicated by the dashed lines in Fig.~\ref{Fig2}E,F (Sec.~IX in SI).

Our observations open up exciting opportunities for exploring strongly-correlated many-body phenomena in bilayer systems, particularly in understanding magnetic exchange instabilities -- one of the important challenges in modern condensed matter physics.
Experimentally, the challenge lies in controllably entering and probing a strongly-interacting regime, while theoretically, the phase diagram for $r_s\simeq 10 - 20$ (as in our experiment), where the electronic system is between a simple Fermi liquid and crystalline states~\cite{zhou2021bilayer,sung2023observation}, is not yet fully understood, with only limited Monte Carlo data. 
In this context, MoS$_2$-homobilayers offer a key advantage as we can naturally access this strongly-interacting regime, while interlayer excitons represent a unique optical probe of pseudospin correlations.

Our observations have close connections with several fundamental  many-body phenomena. 
First, the proposed interlayer electron coherence is closely related to interlayer exciton condensates~\cite{wang2019evidence,ma2021strongly,shi2022bilayer} and, thus, is expected to exhibit superfluid and counterflow responses~\cite{stern2000dissipationless,nguyen2023perfect,spielman2000resonantly,kellogg2004vanishing,kellogg2002observation,spielman2001observation,liu2022crossover,liu2017quantum,li2017excitonic}. 
In this context, our observation in Fig.~\ref{Fig2}F suggests the possibility of superfluidity at temperatures as high as 75$\,$K, even without an applied magnetic field.
Second, while our study reaches a maximum electron density $n$ of about $3\times 10^{12}\,$cm$^{-2}$, further increase in $n$ should eventually melt the electron coherence~\cite{zheng1997exchange}, an expectation supported by the absence of electron tunneling observed at $r_s\simeq 3$~\cite{pisoni2019absence}. 
At densities about an order of magnitude larger than those in our experiment, bilayer superconductivity is expected to emerge~\cite{zhao2023evidence}.
Third, a small twist between the TMD layers breaks the ${\cal C}_3$-rotational symmetry and gives rise to a small direct electron tunneling. 
This tunneling is expected to stabilize the order parameter phase coherence and lead to more coherent rather than stochastic hybridization between interlayer excitons. In addition, the application of an in-plane magnetic field might enable the exploration of the Pokrovsky-Talapov phase transition~\cite{Yang1996Spontaneous} (Sec.~X in SI discusses that even without twisting, electron pair tunneling events can occur but their role is yet to be fully understood).
Fourth, the TMD valley degree of freedom is expected to enrich the phase diagram compared to conventional semiconductors, as the order parameter is likely to have multiple components (Fig.~\ref{Fig4}A and Sec.~IX in SI) -- understanding the structure of spatial order parameter inhomogeneities and their interplay with disorder warrants further theoretical investigation.

Finally, another exciting avenue for future research is to explore the coherence properties of strongly-interacting indirect excitons. Our work demonstrates that these can be substantially influenced by tuning the many-body electron system, potentially enabling novel quantum optics applications. We envision that, similar to the interlayer exciton coupling observed here, electron doping of MoS$_2$-trilayers might lead to the hybridization of quadrupolar excitons~\cite{yu2023observation,leisgang2020giant}, which could have  promising applications for sensing in the terahertz domain and quantum information processing~\cite{yelin2002resonantly}.

\section*{Methods}
\emph{Device fabrication}---2H-stacked bilayer MoS$_2$, hBN, and few-layer-graphite were exfoliated from bulk crystals onto silicon substrates with a 285$\,$nm silicon oxide layer. Bilayer MoS$_2$ flakes were identified according to the reflectance contrast under an optical microscope. The thickness of the hBN flakes was measured by an atomic force microscope. Four graphite/hBN/BL MoS$_2$/hBN/graphite 
heterostructures were fabricated using the dry transfer 
method~\cite{kim2016van}, where electrical contacts were made to the MoS$_2$ and the graphite gates using 10$\,$nm Cr and 100$\,$nm Au deposited via electron beam evaporation. Data from device 1, with top/bottom hBN thicknesses of 19$\,$nm/24$\,$nm, are presented in the main text. Devices 2 and 3 are fabricated with top/bottom hBN thicknesses of 36$\,$nm/38$\,$nm and 32$\,$nm/16$\,$nm, respectively. Device 4 uses thin hBN layers as gate dielectrics, where the top/bottom hBN is 5.4$\,$nm/6.3$\,$nm thick.

\emph{Optical spectroscopy}---Polarization-resolved measurements were conducted in a Bluefors 
dilution refrigerator. All other optical measurements were carried out in a Montana Instruments cryostat (base temperature $T$ = 8$\,$K), using a custom-built 4f confocal setup with a Zeiss (100x, ${\rm NA}=0.75$, ${\rm WD}=4\,$mm) objective. Reflectance spectra were measured using a halogen source (Thorlabs SLS201L) and a spectrometer (Acton SpectroPro 2300i). Electrostatic gating was performed with Keithley 2400 sourcemeters.

\section*{Acknowledgments}
The authors would like to thank I.~Esterlis, C.~Kuhlenkamp, P.~Volkov, A.~Atanasov, E.~Kaxiras, D.~Larson, and A.~Imamoglu for fruitful discussions. 
We acknowledge support from NSF (PHY-2012023 for H.P. and M.D.L.), Center for Ultracold Atoms (an NSF Physics Frontier Center) (PHY-1734011 for H.P. and M.D.L.), AFOSR (FA2386-21-1-4086 for P.K.), and Samsung Electronics (for P.K. and H.P.).
N.L. acknowledges support from the Swiss National Science Foundation (SNF), Project No. P500PT\_206917. 
The work of P.E.D. is sponsored by the Army Research Office and was accomplished under Grant Number W911NF-21-1-0184.
A.A.Z. acknowledges support from Amazon Web Services, Grant Number A50791. K.W. and T.T. acknowledge support from the JSPS KAKENHI (Grant Numbers 21H05233 and 23H02052) and World Premier International Research Center Initiative (WPI), MEXT, Japan.

\section*{Author contributions}

X.L., N.L., A.A.Z, J.S., and J.W. designed and fabricated the devices and performed optical spectroscopy measurements.
X.L., N.L., and P.E.D. performed data analysis and interpreted the results with input from all co-authors.
P.E.D. and V.W. developed the theoretical analysis.
T.T. and K.W. grew the high-quality hBN bulk crystal. 
All co-authors contributed to preparing the manuscript.
All work was supervised by E.D., H.P., P.K., and M.D.L.

\section*{Declarations}

The authors declare no conflicts of interest. Raw data and analysis code are available upon reasonable request.

\bibliography{TMD_lib}

\end{document}


\title{ Supplemental Information for ``Optical signatures of interlayer electron coherence
in a bilayer semiconductor'' }

\author{Xiaoling~Liu}
\thanks{These authors contributed equally to this work.}
\affiliation{\Harvard}
\author{Nadine~Leisgang}
\thanks{These authors contributed equally to this work.}
\affiliation{\Harvard}
\author{Pavel~E.~Dolgirev}
\thanks{These authors contributed equally to this work.}
\affiliation{\Harvard}

\author{Alexander~A.~Zibrov}
\affiliation{\HarvardChem}
\affiliation{\Harvard}

\author{Jiho~Sung}
\affiliation{\HarvardChem}
\affiliation{\Harvard}

\author{Jue~Wang}
\affiliation{\HarvardChem}
\affiliation{\Harvard}

\author{Takashi~Taniguchi}
\affiliation{\JapanFirst}

\author{Kenji~Watanabe}
\affiliation{\JapanSecond}

\author{Valentin~Walther}
\affiliation{\PurduePhys}
\affiliation{\PurdueChem}
\affiliation{\Harvard}

\author{Hongkun~Park}
\affiliation{\HarvardChem}
\affiliation{\Harvard}

\author{Eugene~Demler}
\affiliation{\ETH}

\author{Philip~Kim}
\affiliation{\Harvard}
\affiliation{\HarvardEng}

\author{Mikhail~D.~Lukin}\email[Correspondence to: ]{lukin@physics.harvard.edu}
\affiliation{\Harvard}

\date{\today}

\maketitle
\tableofcontents

\beginsupplement

\begin{figure}[b!]
\centering
\includegraphics[width=180mm]{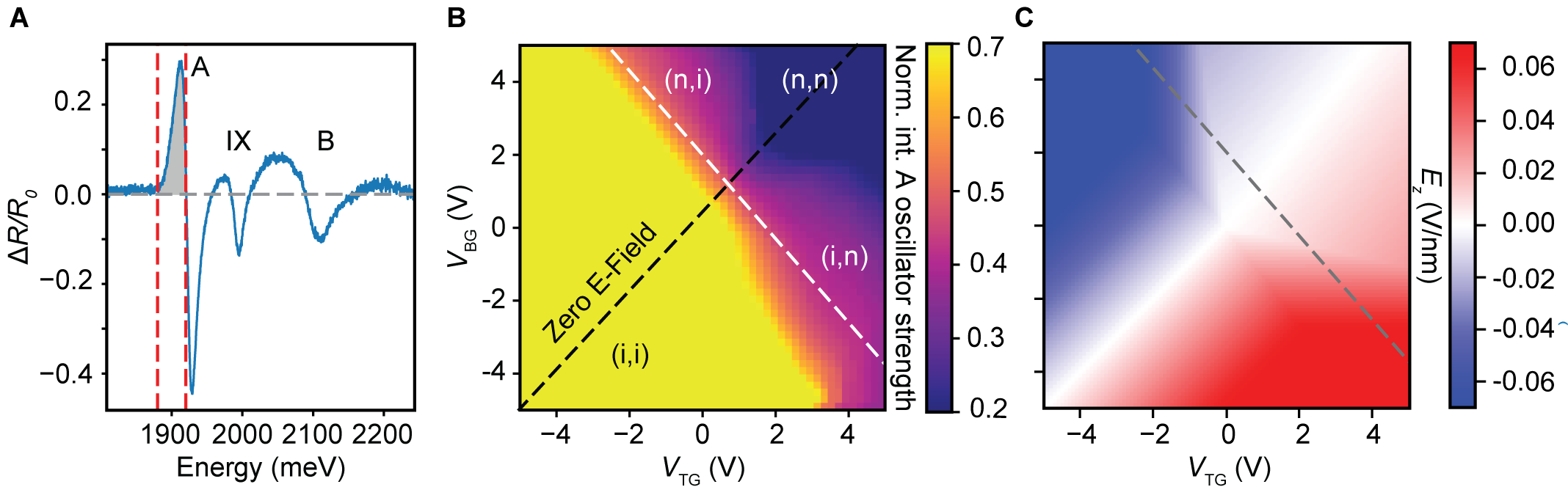} 
\caption{Dual-gated voltage sweep. \textbf{A} Representative differential reflectivity spectrum $\Delta R/R_0$ of the MoS$_2$-homobilayer. The $A$-exciton oscillator strength is represented by integrating the area under the positive part of the $A$-exciton Fano-resonance, indicated by the grey shaded area. \textbf{B} Dual-gated voltage map of the extracted $A$-exciton oscillator strength. Both MoS$_2$ layers are intrinsic (i,i) in the yellow region and electron-doped (n,n) in the blue region. (i,n) and (n,i) label the purple regions where one layer is intrinsic while the other is doped. The black dashed line denotes the simulated $E_z = 0$ line.  \textbf{C} Dual-gated voltage map of the simulated electric field $E_z$. The white dashed line in \textbf{B} and the grey dashed line in \textbf{C} encode the direction of the electric-field sweeps performed in the main text.}
\label{fig::2d_scan}
\end{figure}

\begin{figure}[t!]
\centering
\includegraphics[width=180mm]{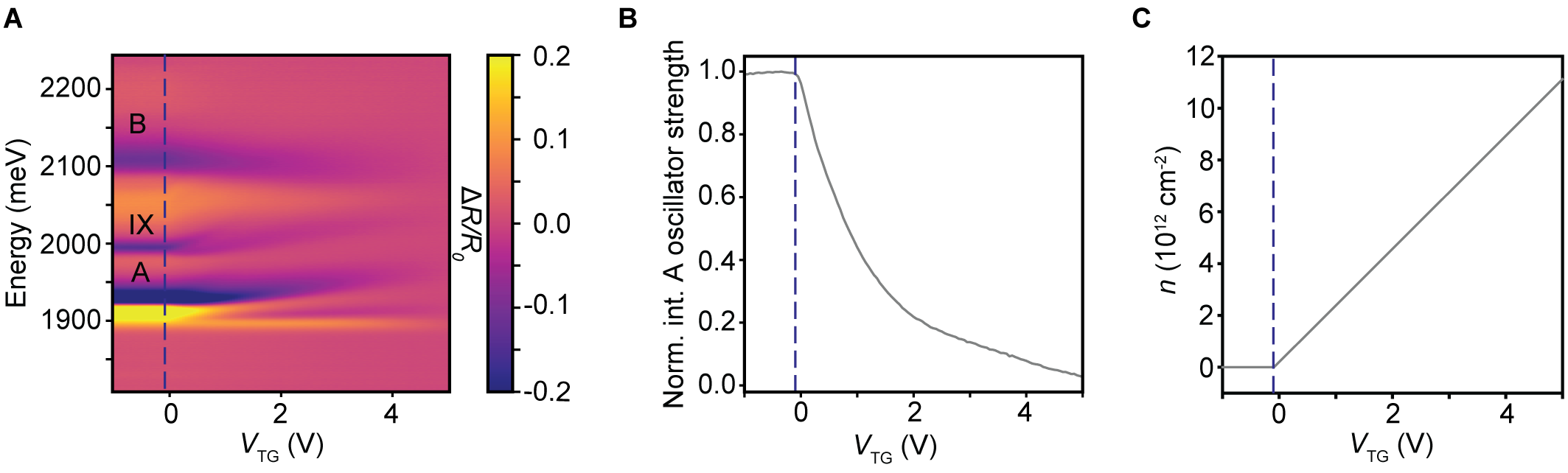} 
\caption{Sweep along the $E_z = 0$ line encodes the evolution of $A$-, IX-, $B$-excitons \textbf{A} and the $A$-exciton oscillator strength \textbf{B}
with doping, simulated in \textbf{C}.  The blue dashed line in  all three panels mark the doping onset. 
}
\label{fig::zeroEline}
\end{figure}

\section{Device electrostatics}
\label{secSI:capacitance model}
\subsection{Dual-gated voltage scan} \label{2dscan}

Doping the sample modifies the $A$-exciton lineshape, splitting its resonance into attractive and repulsive polaron branches. By keeping track of the $A$-exciton, we, therefore, can decipher electrostatic and doping properties of the MoS$_2$-homobilayer.

A representative differential reflectivity spectrum is shown in Fig.~\ref{fig::2d_scan}A, where we identify the $A$-, IX-, and $B$-excitons.
We represent the oscillator strength of the $A$-exciton as the integrated area under the positive part of the respective Fano-resonance peak (shaded grey; red dashed lines denote the boundaries of the integrated energy range).
Integrating over the area at each ($V_{\rm TG}$, $V_{\rm BG}$)-pair and normalizing by the maximum integrated area, we obtain the 2D map in Fig.~\ref{fig::2d_scan}B. 
The brighter/darker region represents a higher/lower $A$-exciton oscillator strength and, thus, a lower/higher doping level. 
Due to the Fermi-level pinning to the  MoS$_2$ conduction band, only electron doping is achievable in our applied gate-voltage range~\cite{singh2019origin,park2023unveiling,gong2014unusual,kang2014computational}.
The direction of the electric-field sweeps described in the main text is illustrated in the 2D map as the white dashed line. 
The black dashed line is the predicted $E_z = 0$ line from our electrostatic simulations in Sec.~\ref{simulation} below.
While our simulations might not be quantitatively accurate due to simplified assumptions about the real system, the electric-field sweeps conducted in the main text always cross the $E_z = 0$ line.
Since we have not observed degeneracy of IX$_{\rm T}$- and IX$_{\rm B}$-excitons at any gate voltage along such sweeps and for a finite density $n\neq 0$, we conclude that these excitons become non-degenerate at $E_z = 0$ when the sample is doped.

\subsection{Electrostatic model}\label{simulation}

We model the electrostatic properties of the sample as follows. 
Simple capacitance equations -- relating the electric potentials at the top gate $V_{\rm TG}$,  bottom gate $V_{\rm BG}$, top TMD layer $\phi_{\rm top}$, and bottom TMD layer $\phi_{\rm bot}$ to the corresponding carrier densities $n_{\rm T}$, $n_{\rm B}$, $n_{\rm top}$, $n_{\rm bot}$ -- are given by:
\begin{subequations}
\begin{align}
    e n_{\rm T} &= \frac{\epsilon_{\rm hBN}}{d_{\rm T}}  (V_{\rm TG} - \phi_{\rm top}),\\
e n_{\rm top} &= \frac{\epsilon_{\rm TMD}}{d_{\rm TMD}}  (\phi_{\rm top} - \phi_{\rm bot}) - \frac{\epsilon_{\rm hBN}}{d_{\rm T}}  (V_{\rm TG} - \phi_{\rm top}),\\
e n_{\rm bot} &= \frac{\epsilon_{\rm hBN}}{d_{\rm B}}  (\phi_{\rm bot} - V_{\rm BG}) - \frac{\epsilon_{\rm TMD}}{d_{\rm TMD}}  (\phi_{\rm top} - \phi_{\rm bot}),\\
e n_{\rm B} &= -\frac{\epsilon_{\rm hBN}}{d_{\rm B}}  (\phi_{\rm bot} - V_{\rm BG}).
\end{align}\label{eqs_capacitance_p1}
\end{subequations}
Here, $\epsilon_{\rm hBN}$ and $\epsilon_{\rm TMD}$ are the permittivities of hBN and MoS$_2$; $d_{\rm T}$, $d_{\rm B}$, and $d_{\rm TMD}$ are the top hBN, bottom hBN, and bilayer MoS$_2$ thicknesses, respectively. Equations~\eqref{eqs_capacitance_p1} are consitent with the charge neutrality condition $n_{\rm T}+n_{\rm top}+n_{\rm bot}+n_{\rm B} = 0$. The TMD sample is grounded and in electro-chemical equilibrium with the corresponding contact, resulting in the conditions:
\begin{align}
\phi_{\rm top}+\frac{\mu_{\rm top}}{e}=\phi_{\rm bot}+\frac{\mu_{\rm bot}}{e}   = 0, 
\end{align}
where $\mu_{\rm top}$ and $\mu_{\rm bot}$ label the chemical potentials of the top and bottom MoS$_2$ layers, respectively.
Disregarding the possibility of hole doping, neglecting the small conduction-band spin-orbit splitting, and assuming the electron state is well captured via a simple Fermi liquid, we relate the TMD densities $n_{\rm top/bot}$ to the corresponding chemical potentials $\mu_{\rm top}$ and $\mu_{\rm bot}$ via
\begin{align}
n_{\rm top/bot}&=
\frac{2 m^*_{\rm e} k_{\rm B} T}{\pi\hbar^2} {\ln}\Big(1+e^{(\mu_{\rm top/bot}-\mu_0)/k_{\rm B} T}\Big) ,
\label{eqn:all-lines}
\end{align}
where $m^*_{\rm e}$ is the effective electron mass and $\mu_0$ is a fitting parameter chosen to ensure that the simulated density onset matches our measurements -- see Fig.~\ref{fig::zeroEline}.

\newpage

\section{Reproducibility of the observations}
\label{sec_SI_reproducibility}

\begin{figure}[t!]
\centering
\includegraphics[width=180mm]{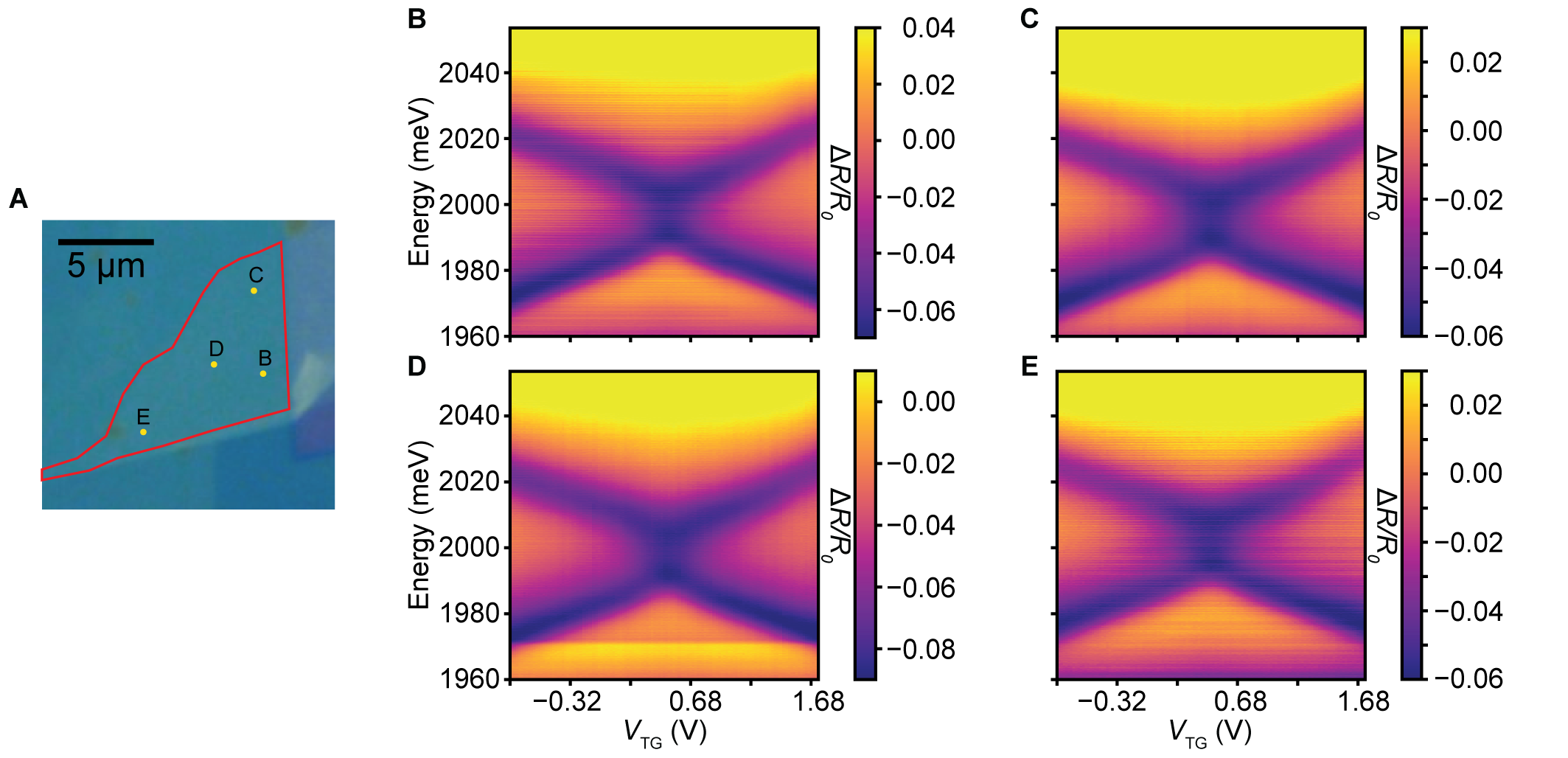} 
\caption{Reproducibility of the stochastic anti-crossing within the same device.
 \textbf{A} Microscope image of the main device~1; the red line outlines the region containing the dual-gated MoS$_2$-homobilayer.
 \textbf{B}-\textbf{E} Electric-field sweeps at four representative optical spots marked in \textbf{A} confirming that the stochastic anti-crossing is robustly present across the sample.}
\label{fig::device1_spots}
\end{figure}

\emph{Reproducibility within the same device.---}Figure~\ref{fig::device1_spots} shows that the stochastic anti-crossing is robustly present across the entire spatial extend of the main device (device~1), confirming that this effect is highly reproducible within the same sample.

\begin{figure}[t!]
\centering
\includegraphics[width=180mm]{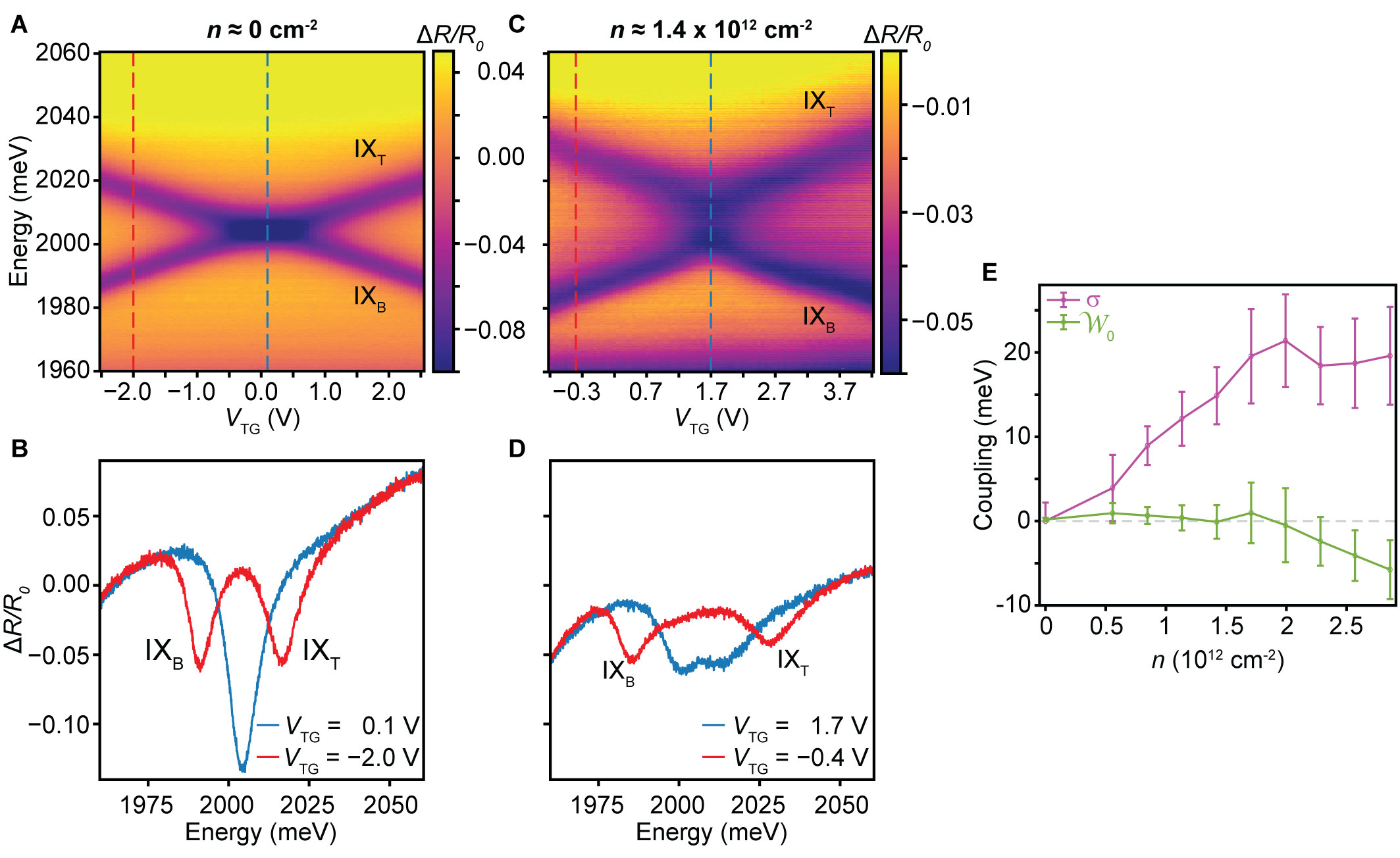} 
\caption{The stochastic anti-crossing in device 2. \textbf{A} Electric-field sweep in the intrinsic region ($n \approx 0$), demonstrating a simple interlayer exciton crossing. \textbf{B} Two linecuts at $E_z = 0$ and $E_z\neq 0$, marked by white dashed lines in \textbf{A}, illustrate amplitude doubling at the degeneracy point $E_z = 0$.
\textbf{C,D} Similar to \textbf{A,B}, but for the doped system, revealing a transition from a simple crossing to a stochastic crossing. In the doped case, the linecut at $E_z = 0$ no longer exhibits amplitude doubling.
\textbf{E} Evolution of the mean coupling ${\cal W}_0$ and the variance $\sigma$ with electron density, obtained by fitting the 2D reflectance maps (see Sec.~\ref{sec_SI_data} below), showing quantitative agreement with the results presented in Fig.~2 of the main text.}
\label{fig::device2}
\end{figure}

\emph{Reproducibility in other devices.---}We fabricated two additional devices (device~2 and device~3) with roughly (but not exactly) the same geometry. 
Figure~\ref{fig::device2} shows the stochastic anti-crossing measurements in device~2 (device~3 is not shown as a similar effect was observed there), confirming not only the reproducibility of this phenomenon in another device but also the robustness of our data analysis. The extracted values for ${\cal W}_0$ and $\sigma$ agree quantitatively with those for device~1 in Fig.~2 of the main text.

\section{Evidence for the Coulomb origin of the stochastic anti-crossing}\label{thinBN}

\begin{figure}[t!]
\centering
\includegraphics[width=180mm]{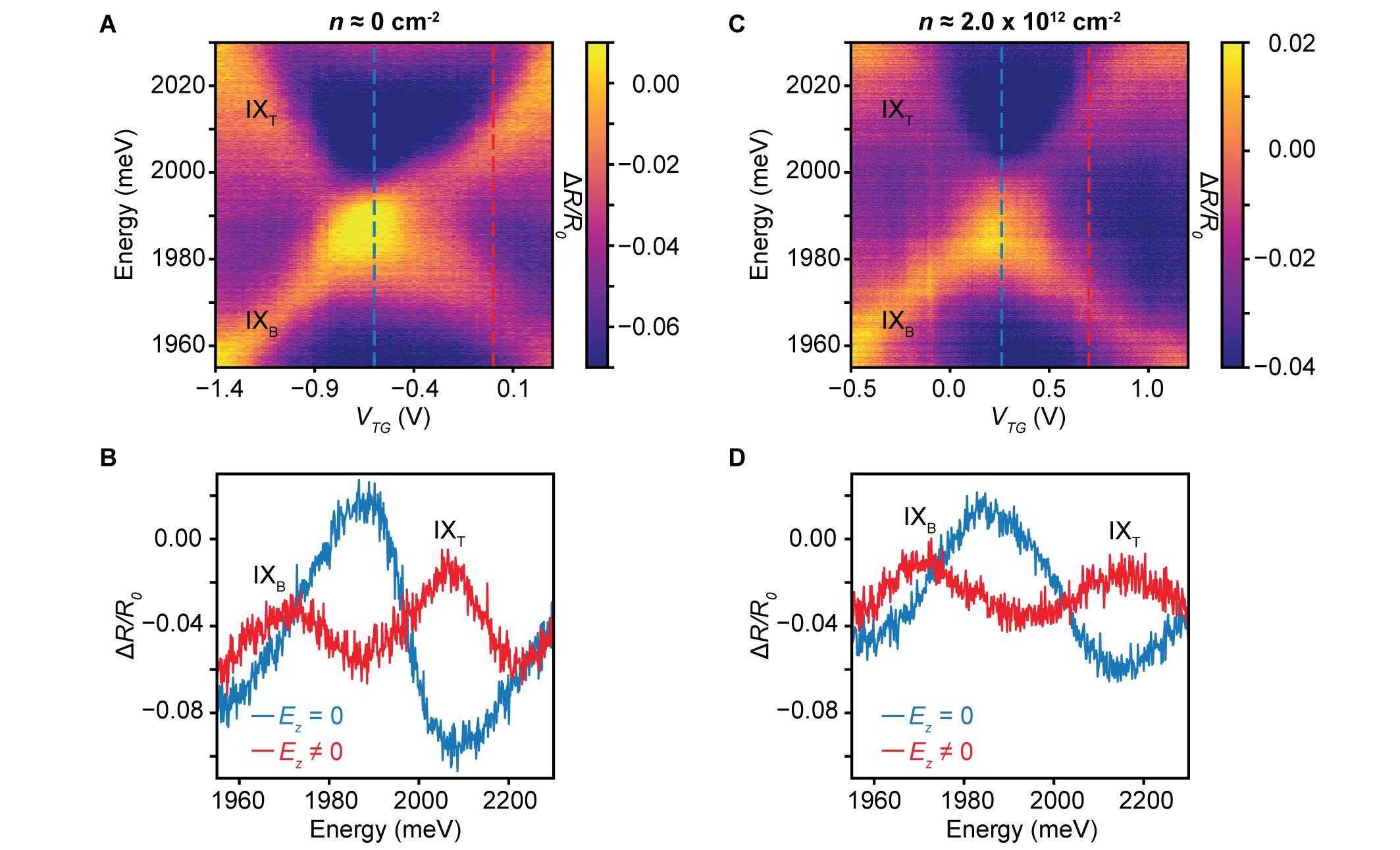} 
\caption{Coulomb origin of the stochastic anti-crossing.
\textbf{A} Electric-field sweeps for the undoped case in device~4, with thin hBNs as gate dielectrics, show the crossing of two interlayer excitons at the degeneracy point $E_z = 0$ ($T = 4\,$K). Linecuts (white dashed lines in \textbf{A}) at $E_z = 0$ and $E_z \neq 0$ in \textbf{B} reveal amplitude doubling at $E_z = 0$. In device 4, interlayer excitons exhibit peaks, rather than dips, in reflectance spectra, attributed to interference with the thin-hBN background reflectivity.
 \textbf{C,D} Doping this thin-hBN sample does not disrupt the crossing and degeneracy of the two interlayer excitons at $E_z = 0$.
}
\label{fig::thinBN}
\end{figure}

To investigate the Coulomb origin of the stochastic anti-crossing, we fabricated another device (designated as device 4) using thin hBNs as gate dielectrics, as described in the Methods section. This new device features a substantially different surrounding Coulomb environment compared to devices~1-3 with thick hBNs.
Additionally, the thin hBN dielectric enables screening of Coulomb interactions in the MoS$_2$-homobilayer from the graphite gates due to the short distance between the sample and the gates~\cite{tebbe2023tailoring}. Figure~\ref{fig::thinBN} demonstrates that device~4 no longer exhibits the stochastic anti-crossing due to this screening effect, confirming the Coulomb origin of the observed phenomenon.

\section{Temperature effects}
\label{sec:temp}

\begin{figure}[t!]
\centering
\includegraphics[width=180mm]{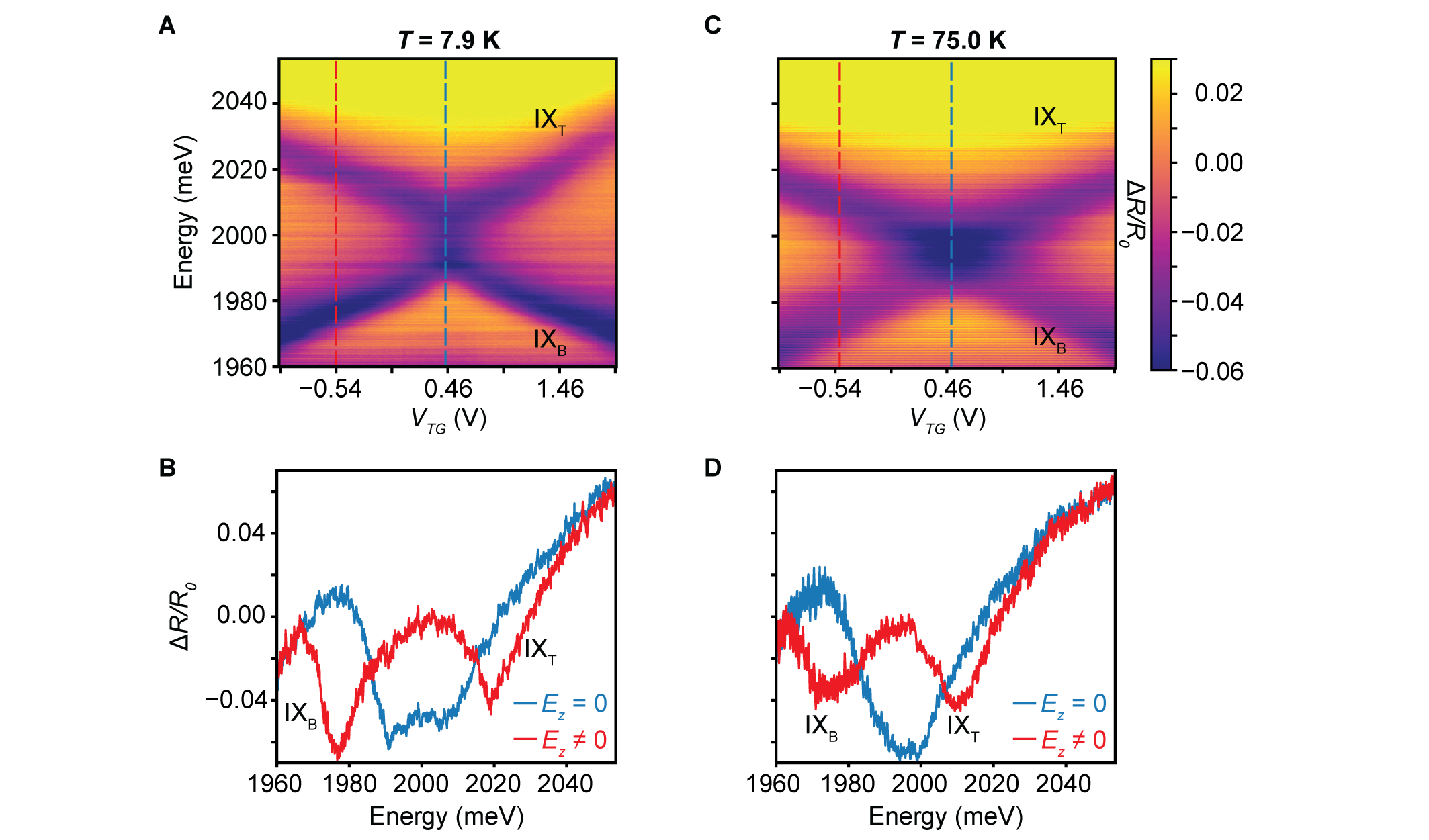} 
\caption{Electric-field sweeps at two representative temperatures at $n\approx 1.3\times 10^{12}\,$cm$^{-2}$. The stochastic DC Stack effect, clearly observed at $T = 7.9\,$K (panels \textbf{A} and \textbf{B}), disappears at $T = 75.0\,$K (panels \textbf{C} and \textbf{D}) and turns into a simple crossing, as further evidenced by the amplitude doubling at the degeneracy point $E_z = 0$ in \textbf{D} (see also Fig.~\ref{fig::device2}B).
}
\label{fig::lowT_vs_highT}
\end{figure}

Figure~\ref{fig::lowT_vs_highT} presents raw data from stochastic anti-crossing measurements at $n\approx 1.3\times 10^{12}\,$cm$^{-2}$, comparing low ($T = 7.9\,$K) and high ($T = 75\,$K) temperatures. 
At $T = 7.9\,$K, stochastic hybridization is evident, while at $T = 75\,$K, it transitions to a simple crossing, characterized by amplitude doubling at the degeneracy point $E_z = 0$ in Fig.~\ref{fig::lowT_vs_highT}D.
While at higher temperatures processes such as exciton scattering off phonons become progressively more important and result in excitonic line broadening,  this broadening alone does not fully account for the observed relative amplitude change.
For this reason, and following the main text, we interpret our temperature-dependent measurements as evidence of order parameter melting, see also Sec.~\ref{sec_SI_main_theory}.

\section{Optical spot size effects}
\label{sec_SI_spot_size}

In the main text, we proposed that the static stochastic variance $\sigma$ originates from an order parameter -- in the form of interlayer electron coherence -- of the many-body electron system. We attribute the origin of this static stochasticity $\sigma$ to immobile spatial order parameter fluctuations. 
In principle, if the phase coherence length $\xi$ is comparable to the optical spot size $R$, this order parameter could also contribute to the mean coupling ${\cal W}_0$. 
Furthermore, it is tempting to attribute the observed trends of ${\cal W}_0$ in Fig.~2E,F to this long-range scenario. 
We remark that if the optical spot size can be made small, such that $\xi \gtrsim R$, the interlayer exciton hybridization should be coherent without strong static stochastic features, implying that we expect $\xi \lesssim R$ in the experiment.

\begin{figure}[t!]
\centering
\includegraphics[width=180mm]{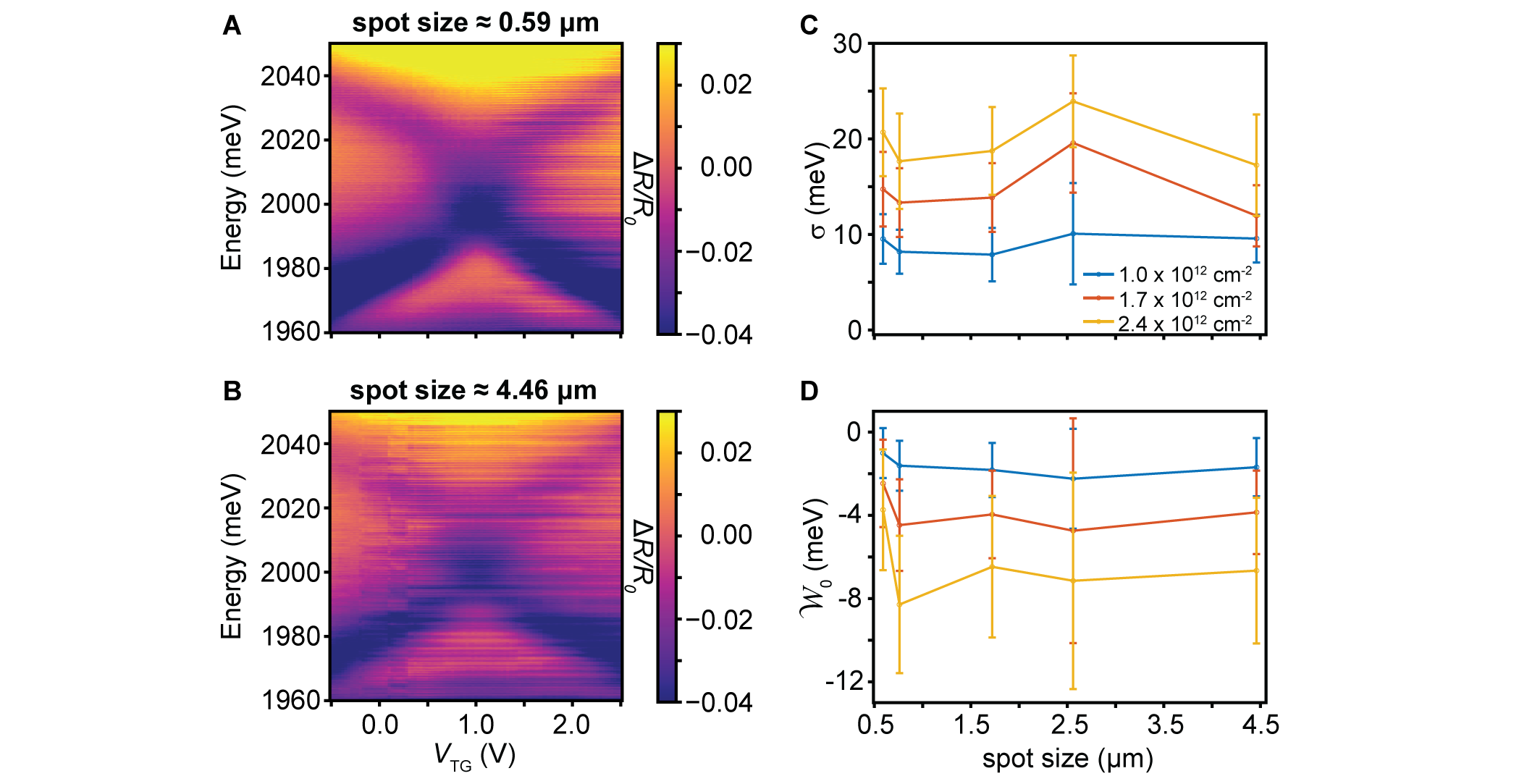} 
\caption{Optical spot size effects. \textbf{A} and \textbf{B} show electric-field sweeps at $n \approx 2.33\times 10^{12}\,$cm$^{-2}$ for a focused (\textbf{A}) and defocused (\textbf{B}) optical spot size. \textbf{C} and \textbf{D} depict fitted values of $\sigma$ and ${\cal W}_0$, respectively, which do not exhibit a clear dependence on the optical spot size.
}
\label{fig::spot_size}
\end{figure}

One way to test this long-range scenario is to study how the stochastic anti-crossing varies with the optical spot size, as we expect that the order-parameter contribution to the mean coupling behaves as $\delta{\cal W}_0\propto \xi/R$. 
Figure~\ref{fig::spot_size}A,B shows that upon increasing the optical spot size by about an order of magnitude, extracted by fitting the camera image of a laser spot with a 2D Gaussian, both the stochastic variance $\sigma$ and the mean coupling ${\cal W}_0$ roughly remain intact. These data are, therefore, consistent with the picture where the phase coherence length is appreciably smaller than the smallest optical spot size of about $0.6\,\mu$m, so that the order parameter can contribute to $\sigma$ but not to ${\cal W}_0$.

\clearpage
\newpage

\section{Static stochastic avoided crossing}
\label{sec_SI_osc_model}

In this section, we detail our phenomenological modeling of the stochastic crossing representing our measurements of the DC Stark effect at a finite electron density. 

The two interlayer excitons with opposite out-of-plane dipole moments can be modeled by a simple model of two coupled harmonic oscillators (here we set the zero of energy to be the interlayer exciton energy at the degeneracy point $E_z = 0$):
\begin{align}
    i\partial_t X_{\rm T} = \omega_{\rm T} X_{\rm T} - i \gamma_{\rm T} X_{\rm T} + {\cal W} X_{\rm B}  - d_{\rm T} {\cal E}_d(t),\quad
    i \partial_t X_{\rm B} = \omega_{\rm B} X_{\rm B} -i \gamma_{\rm B} X_{\rm B} + {\cal W} X_{\rm T} - d_{\rm B} {\cal E}_d(t). \label{eqn: coupled_osc}
\end{align}
Here $\omega_{\rm T/B} = \pm d_z E_z$ encodes the bare excitonic energies that linearly shift with the out-of-plane electric field $E_z$; $d_z$ is the out-of-plane dipole moment; $\gamma_{\rm T/B}$ is the total linewidth of the corresponding excitonic resonance; ${\cal W}$ is the coupling strength between the two excitonic branches; $d_{\rm T/B}$ is the transition dipole moment (assumed to be real in our modeling) encoding the response to the probe field ${\cal E}_d$. The observable we are interested in is the imaginary part of the susceptibility ${\rm Im}[\chi(\omega)]$ defined as ${\cal P}(\omega) = \chi(\omega) {\cal E}_d(\omega)$, where ${{\cal P}(\omega)}= d_{\rm T} X_{\rm T}(\omega) + d_{\rm B} X_{\rm B}(\omega)$ is the TMD excitonic polarization.

\subsection{Modeling the fluctuating coupling }

We consider two scenarios for the fluctuating coupling ${\cal W}$:
\begin{itemize}
    \item \textbf{Static fluctuations model}: ${\cal W}$ is time-independent but the measured signal represents an average over the distribution with
    \begin{align}
        \langle {\cal W}\rangle = {\cal W}_0,\qquad  \delta {\cal W} = ({\cal W} - {\cal W}_0) \in [-\sigma, \sigma]. 
        \label{eqn:homog}
    \end{align}
    \item  \textbf{Dynamic white-noise averaging}: 
    \begin{align}
        \langle {\cal W}(t) \rangle = {\cal W}_0, \qquad   \langle \delta {\cal W}(t)\delta {\cal W}(t') \rangle =  \gamma \delta(t - t').\label{eqn:dyn}
    \end{align}
\end{itemize}
We note that in the former limit, the coupling ${\cal W}(t)$ can, in principle, be time-dependent, but its dynamics should occur on timescales much longer than the exciton dynamics, set by the splitting $\Delta = 2d_z E_z$ as well as by the decay rates $\gamma_{\rm T}$ and $\gamma_{\rm B}$. The latter scenario corresponds to the opposite limit where the dynamics of ${\cal W}(t)$ are much faster than any other relevant timescales, allowing for the approximation of the variable ${\cal W}(t)$ as Markovian.

\subsection{Analysis of the static fluctuations scenario}

\begin{figure}[t!]
\centering
\includegraphics[width=180mm]{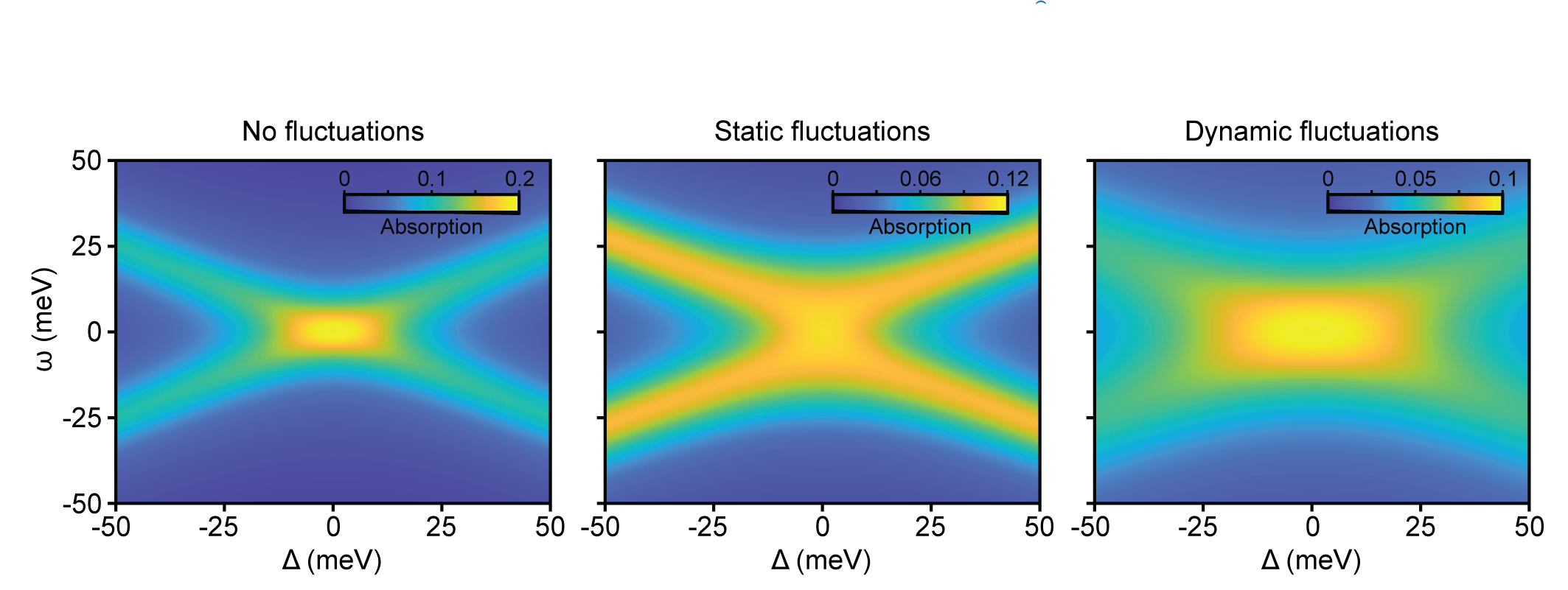} 
\caption{ Absorption maps ${\rm Im}[\chi]$ for the two scenarios of fluctuating coupling.  The left panel depicts the case of uncoupled interlayer excitons. The middle panel illustrates the static stochastic scenario with ${\cal W}_0 = 0$ and $\sigma = 20\,$meV. The right panel shows the dynamic scenario with $\gamma = 20\,$meV. Parameters used: $d_{\rm T}=d_{\rm B}$, $\gamma_{\rm T} = \gamma_{\rm B} = 10\,$meV. }
\label{fig::Stochastic coupling}
\end{figure}

For a given ${\cal W}$,  the response function is given by:
\begin{align}
    \chi_{\cal W}(\omega) = \frac{-1}{(\omega - \Delta/2 + i\gamma_{\rm T})(\omega + \Delta/2 + i\gamma_{\rm B}) - {\cal W}^2} \begin{bmatrix}
        d_{\rm T}\\
        d_{\rm B}
    \end{bmatrix}^{\rm T}
    \begin{bmatrix}
        \omega + \Delta/2 + i\gamma_{\rm B} & {\cal W}\\
        {\cal W} & \omega - \Delta/2 + i\gamma_{\rm T} 
    \end{bmatrix}\begin{bmatrix}
        d_{\rm T}\\
        d_{\rm B}
    \end{bmatrix}. \label{eqn:chi}
\end{align}
We are interested in the average response $\langle  \chi_{\cal W}(\omega) \rangle$ over the distribution in Eq.~\eqref{eqn:homog}. This computation can be done analytically, and it boils down to evaluating the following integrals:
\begin{align}
    {I}_1  = \frac{1}{2 C}  \int_{-\sigma}^{\sigma} \frac{d {\cal W}}{2 \sigma} \Big[ \frac{1}{{\cal W} + {\cal W}_0 - C} - \frac{1}{{\cal W} + {\cal W}_0 + C} \Big],\quad
    {I}_2  = \frac{1}{2 }  \int_{-\sigma}^{\sigma} \frac{d{\cal W}}{2 \sigma}  \Big[ \frac{1}{{\cal W} + {\cal W}_0 - C} + \frac{1}{{\cal W} + {\cal W}_0 + C} \Big], \label{eqn:I_exprs}
\end{align}
where $C^2 = (\omega - \Delta/2 + i\gamma_{\rm T})(\omega + \Delta/2 + i\gamma_{\rm B})$.
These integrals can be computed using $\displaystyle\int_{-\sigma}^{\sigma}   \frac{d{\cal W}}{{\cal W} +A} = \log[A + \sigma] - \log[A - \sigma]$. We verified that numerical averaging of Eq.~\eqref{eqn:chi} matches the results obtained from the analytical approach. The resulting response function is shown in Fig.~\ref{fig::Stochastic coupling} and well represents the measured data. We remark that we use here the uniform distribution in Eq.~\eqref{eqn:homog} only for numerical convenience, and any other reasonable distribution (e.g., Gaussian) could be used instead.

\subsection{Analysis of the dynamic fluctuations scenario}

Dynamic modeling requires additional considerations as the fluctuating coupling term, $\delta {\cal W}(t)$, introduces multiplicative noise. Therefore, it may be essential to consider parametric processes. To see this point explicitly, we write the equations of motion in the frequency domain:
\begin{align}
    \hat{\cal G}_{0,\omega} \begin{bmatrix}
        X_{\rm T}(\omega)\\
        X_{\rm B}(\omega)
    \end{bmatrix} - \int_{-\infty}^\infty \frac{d\omega'}{2\pi} \begin{bmatrix}
        0 & \delta {\cal W}_{\omega - \omega'}\\
        \delta {\cal W}_{\omega - \omega'} & 0
    \end{bmatrix} \begin{bmatrix}
        X_{\rm T}(\omega')\\
        X_{\rm B}(\omega')
    \end{bmatrix} = - {\cal E}_d(\omega )\begin{bmatrix}
        d_{\rm T}\\
        d_{\rm B}
        \end{bmatrix}, 
        \quad  \hat{\cal G}_{0,\omega} = \begin{bmatrix}
        \omega - \Delta/2 + i\gamma_{\rm T} & -{\cal W}_0\\
        -{\cal W}_0 & \omega + \Delta/2 + i\gamma_{\rm B}
    \end{bmatrix}. \notag
\end{align}
The second term 
encodes the mentioned frequency mixing due to the dynamics of $\delta {\cal W}(t)$, which can cause a parametric instability provided $\delta {\cal W}(t)$ contains Fourier harmonics at frequencies commensurate with the bare Rabi oscillations. In our case, we assume a broadband white noise
   $\langle \delta {\cal W}_{\omega} \delta {\cal W}_{\omega'} \rangle =2\pi \gamma\delta(\omega + \omega')$,
which contains all possible frequency harmonics, including the commensurate ones.

There are various ways to approach the problem at hand, including the direct numerical sampling over the dynamical noise,
and among them, the most efficient appears to be the non-equilibrium Green's function technique developed in Ref.~\cite{dolgirev2020non}. If one is interested solely in the retarded susceptibility, the self-consistent Born approximation turns out to be exact so that the solution can be written as:
\begin{align}
    \hat{\Sigma }^R_\omega = -i\frac{\gamma}{2}\hat{1}\Rightarrow \begin{bmatrix}
        X_{\rm T}(\omega)\\
        X_{\rm B}(\omega)
    \end{bmatrix} = -{\cal E}_d(\omega) [ \hat{\cal G}_{0,\omega}  - \hat{\Sigma }^R_\omega ]^{-1} \begin{bmatrix}
        d_{\rm T}\\
        d_{\rm B}
    \end{bmatrix}.
\end{align}
In other words, the effects of dynamical fluctuations are fully captured via the substitution:
\begin{align}
    \gamma_{\rm T} \to \gamma_{\rm T} + \gamma/2, \qquad  \gamma_{\rm B} \to \gamma_{\rm B} + \gamma/2.
\end{align}
The dynamical fluctuations model can, thus, be understood as a microscopic model of a pure dephasing channel.
The resulting response function, plotted in Fig.~\ref{fig::Stochastic coupling} (right panel), clearly does not capture the observations.

\section{Development of a nonzero mean coupling ${\cal W}_0\neq 0$}
\label{sec:finite W0}

In this section, we present experimental evidence that a nonzero mean coupling ${\cal W}_0\neq 0$ develops at a finite electron density and low temperatures. We also discuss the implications of ${\cal W}_0\neq 0$ for the nature of interlayer exciton hybridization and for rigorous data processing.

\subsection{Experimental evidence}

\begin{figure}[t!]
\centering
\includegraphics[width=180mm]{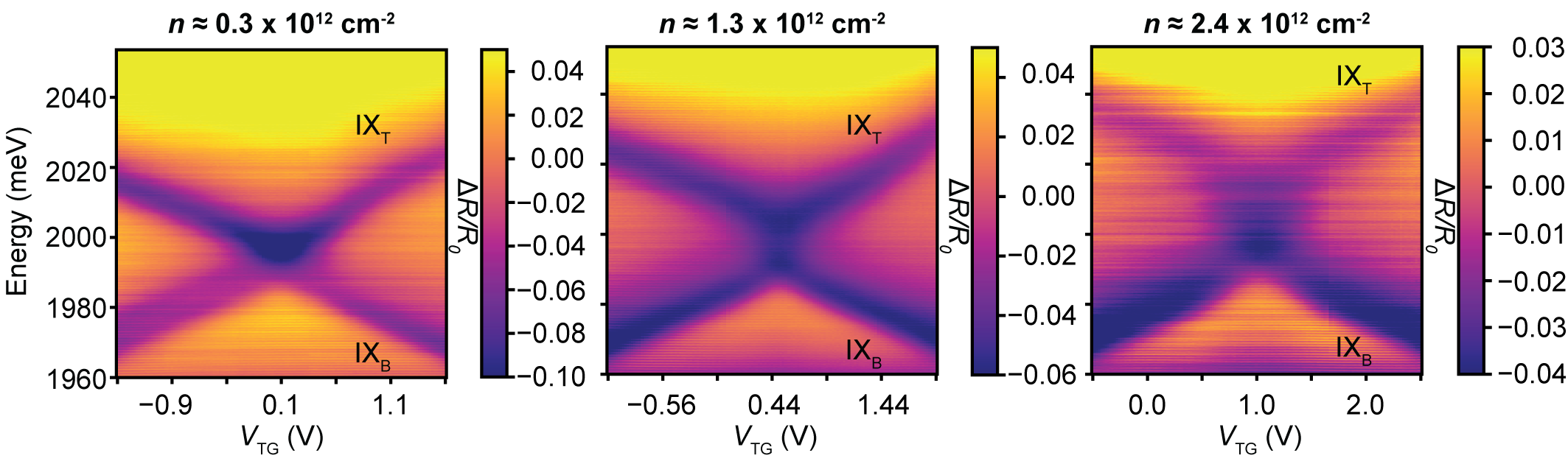} 
\caption{Electric-field sweeps at three representative carrier densities at $T$ = 8~K: $n\approx 0.34\times 10^{12}\,$cm$^{-2}$ (left), $n\approx 1.26\times 10^{12}\,$cm$^{-2}$ (middle), and $n\approx 2.33\times 10^{12}\,$cm$^{-2}$ (right). The evolution from the upper exciton branch initially being stronger than the lower branch, then becoming equal, and eventually becoming weaker, indicates the development of a non-zero ${\cal W}_0\neq 0$.}
\label{fig::3dopingSweeps}
\end{figure}

The presence of a nonzero coupling ${\cal W}_0 \neq 0$ leads to an asymmetry between the upper and lower excitonic branches (see also Fig.~2 of the main text). This effect can be understood as the onset of superradiant and subradiant states of the two coupled dipoles -- see Sec.~\ref{sec_SI_osc_model}. Figure~\ref{fig::3dopingSweeps} shows the evolution of the stochastic anti-crossing with electron density: at low dopings (left panel), the upper branch exhibits slightly stronger oscillator strength; at intermediate dopings (middle panel), the strengths of the two exciton branches are comparable; at high dopings (right panel), the lower branch is brighter, while the upper branch weakens and eventually becomes barely observable. These raw data unambiguously indicate ${\cal W}_0\neq 0$ (see also Fig.~2 of the main text).

\begin{figure}[t!]
\centering
\includegraphics[width=180mm]{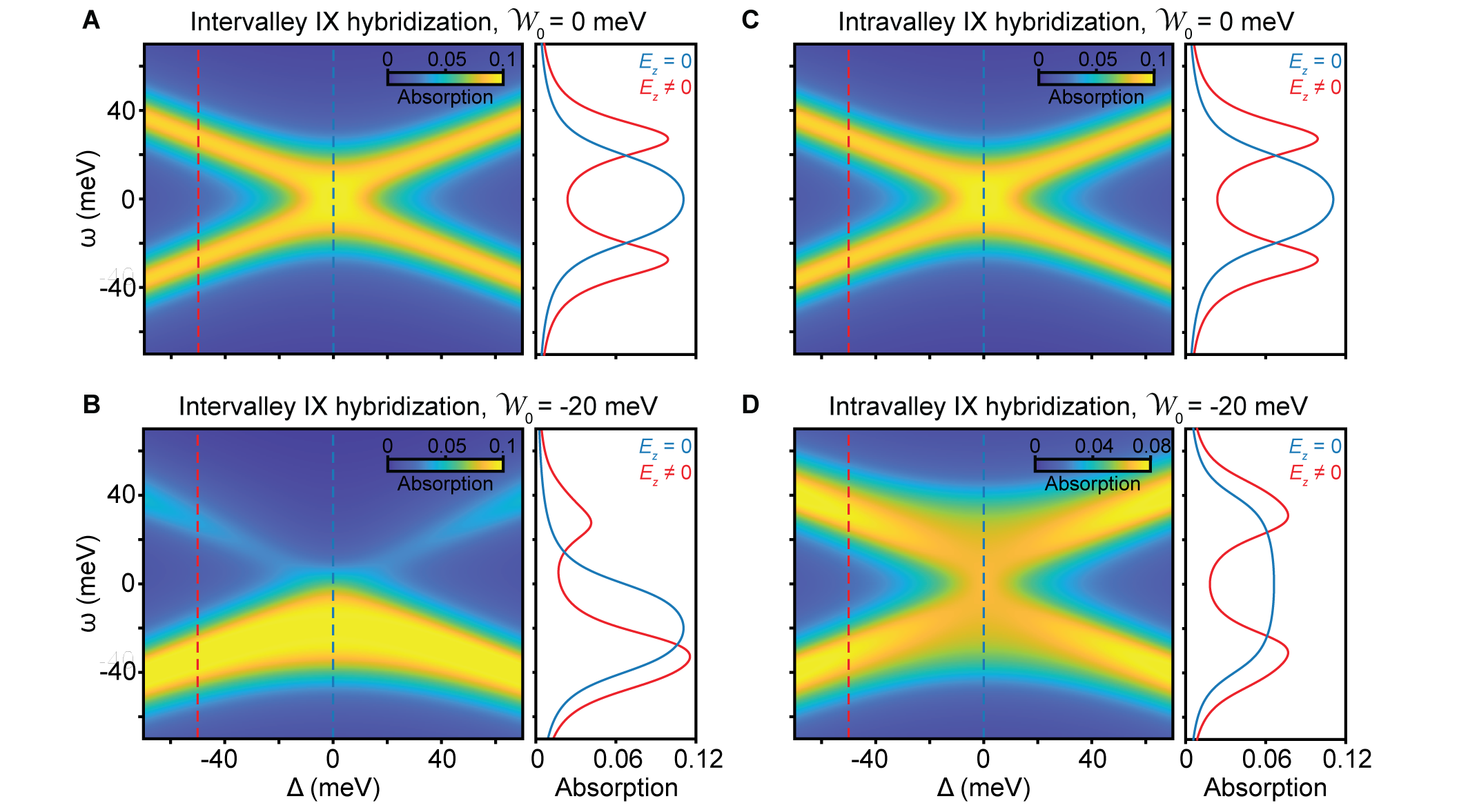} 
\caption{
Intravalley vs intervalley exciton hybridization.
\textbf{A,B} Assuming the case of intervalley hybridization, the simulated stochastic anti-crossing for ${\cal W}_0 = 0$ \textbf{A} and for ${\cal W}_0 = -20\,$meV \textbf{B}  illustrates that a nonzero ${\cal W}_0\neq 0$ leads to an intensity asymmetry between the upper and lower exciton branches. Such asymmetry emerges due to an interference effect between the two intervalley excitons with the same AQNs. In contrast, for the case of intravalley hybridization~\textbf{C,D}, the stochastic anti-crossing shows no such asymmetry even for ${\cal W}_0\neq 0$ because intravalley interlayer excitons have opposite AQNs and, thus, cannot interfere. Here we fixed $\sigma = 20\,$meV.
}
\label{fig::intervalley_hybrid}
\end{figure}

\subsection{ Hybridization of interlayer excitons with the same AQNs}

One immediate implication of a nonzero ${\cal W}_0\neq 0$ is that interlayer excitons with the same AQNs hybridize with each other. This implies that excitons in the opposite valleys hybridize with each other, as excitons within the same valley have opposite AQNs and cannot interfere with each other (i.e., subradiant and superradiant states cannot form in this case).
This conclusion is further illustrated by a simple simulation as in Sec.~\ref{sec_SI_osc_model} -- see Fig.~\ref{fig::intervalley_hybrid}. 
Our analysis, however, does not rule out the possibility that excitons within the same valley can also hybridize (Fig.~\ref{fig::intervalley_hybrid}), which might be relevant in case ${\cal W}_0$ and $\sigma$ have different origins, as further discussed in the main text and in Secs.~\ref{sec_SI_main_theory} and~\ref{sec_SI_FS}.

\subsection{Additional considerations for data analysis}

Below we analyze the experimental data using the static fluctuations model, encoded in Eqs.~\eqref{eqn: coupled_osc} and~\eqref{eqn:homog}, and here we discuss two additional considerations needed for robust data processing:
\begin{itemize}
    \item \underline{Effects of $\sigma$ and $\gamma_{\rm T/B}$:} We first note that for $\Delta = 0$, distinguishing the effects of $\sigma$ from linewidth broadenings $\gamma_{\rm T}$ and $\gamma_{\rm B}$ can be challenging. However, when $|\Delta|\gtrsim \sigma$, the excitonic peak linewidths are primarily determined by $\gamma_{\rm T}$ and $\gamma_{\rm B}$. 
     Therefore, in our data analysis, we simultaneously consider the entire 2D reflectivity map to unambiguously determine $\sigma$; we also fix $\gamma_{\rm T} = \gamma_{\rm B} = \gamma$, where $\gamma$ is assumed to be independent of the applied electric field. 
    \item \underline{Effects of a finite ${\cal W}_0 \neq 0$ and $d_{\rm T}\neq d_{\rm B}$:} We also note that two additional factors can manifest in the measured reflectivity maps as ${\cal W}_0\neq 0$. One of these is interference effects from the background, which we found to be unimportant -- see Sec.~\ref{sec_SI_reproducibility} and Fig.~\ref{fig::device2}E. The other factor arises from the hybridization between the interlayer excitons with the $B$-excitons~\cite{sponfeldner2022capacitively}, resulting from hole tunneling.
    The $B$-excitons favor the higher-energy interlayer exciton to be brighter, an effect that can be captured via $d_{\rm T}\neq d_{\rm B}$ but difficult to unambiguously disentangle from ${\cal W}_0\neq 0$. On the other hand, this asymmetry effect due to the $B$-excitons is expected to be weak, as further supported by $|{\cal W}_0|\lesssim 2\,$meV for the intrinsic region ($n = 0$). To avoid overfitting, we, thus, set $d_{\rm T} = d_{\rm B} = d$, where $d$ is assumed to be electric field independent. This assumption is intuitive because the two interlayer excitons should be degenerate at $\Delta = 0$, but it can lead to a small systematic error in determining ${\cal W}_0$.
\end{itemize}

\section{Data processing}
\label{sec_SI_data}

When one considers reflectivity properties of a TMD sample, it is natural to separate the background contribution $R_{\rm bg}(\omega)$ from the excitonic resonances~\cite{smolenski2021signatures,scuri2018large,zhou2017probing}:
\begin{align}
    R(\omega) \approx R_{\rm bg}(\omega) - \text{Im}[ e^{i\varphi(\omega) } (\chi_{\rm IX}(\omega) + \chi_{\rm A}(\omega))],\label{eqn:R}
\end{align}
where $\varphi(\omega)$ encodes the effects due to the interference between the background and excitonic parts. In analyzing the measured data, we employ Eqs.~\eqref{eqn:chi} and~\eqref{eqn:I_exprs} to fit the interlayer exciton contribution and 
\begin{align}
    \chi_{\rm A}(\omega) = -\frac{d_{\rm A}^2}{\omega - \omega_{\rm A} + i\gamma_{\rm A}}\label{eqn:chiA}
\end{align}
to fit the $A$-exciton contribution.
While we are primarily interested in the IX-properties, we consider a rather large fitting energy range (about 100$\,$meV) so that the $A$-exciton part cannot be fully disregarded -- see Fig.~\ref{fig::3dopingSweeps}. We let $R_{\rm bg}(\omega)\approx R_0 + R_1 (\omega-2000\,$meV$)$ to be linearly dependent on $\omega$ for $|\omega-2000\,$meV$| \ll 2000\,$meV (the choice of 2000$\,$meV is close to the interlayer exciton energy at $E_z = 0$). In principle, $\varphi(\omega)$ can also depend on $\omega$, but we verified that approximating $\varphi(\omega)$ as constant, denoted as $\varphi_0$, provides good fits to the data.

In our analysis, we fit the normalized reflectance data  ${\cal S}(\omega) = (\textit{R}_{\text{TMD}}-\textit{R}_{\text{no-TMD}})/ \textit{R}_{\text{no-TMD}}$, using a similar fitting form as in Eq.~\eqref{eqn:R}. Here, $R_{\text{no-TMD}}$ represents the measured reflectance from the graphite/hBN/hBN/graphite heterostructure, i.e., from the full stack but without the TMD part -- it should reasonably well approximate $R_{\rm bg}$ in Eq.~\eqref{eqn:R}.

\begin{figure}[t!]
\centering
\includegraphics[width=180mm]{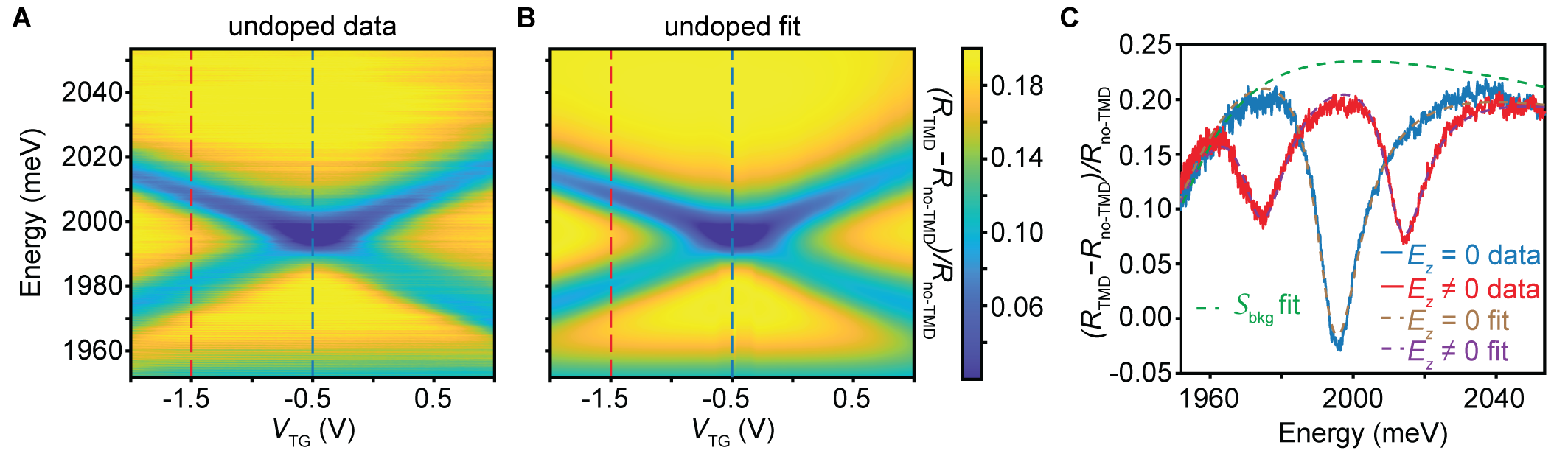} 
\caption{
Data processing when the sample is undoped. \textbf{A} and \textbf{B} show a back-to-back comparison of the measured (\textbf{A}) and fitted (\textbf{B}) electric-field sweeps, demonstrating excellent quantitative agreement. \textbf{C} This agreement is further corroborated by individual linecuts at $E_z = 0$ and $E_z \neq 0$. The green dashed line represents the fitted background, which includes the $A$-exciton contribution.}
\label{fig::intrinsic_fit}
\end{figure}

\subsection{Fitting the intrinsic data}
\label{subsec:intrinsic}

We use the intrinsic data to fix the three background parameters (${\cal S}_0$, ${\cal S}_1$, and $\varphi_0$), which are assumed to be independent of applied voltages and, thus, electron doping. 
Here, each interlayer exciton is modeled by a Lorentzian as in Eq.~\eqref{eqn:chiA}, with three parameters ($\omega_{\rm T/B}$, $\gamma_{\rm T/B}$, and $d_{\rm T/B}$) that can vary with $E_z$ ($V_{\rm TG}$).
The $A$-exciton parameters ($d_A$, $\omega_A$, and $\gamma_A$) can depend on the electron density but not on the electric field $E_z$, acting as a correction to the background $R_{\rm bg}$. Figure~\ref{fig::intrinsic_fit} demonstrates that this fitting approach accurately captures the intrinsic data, allowing us to confidently estimate the background parameters ${\cal S}_0$, ${\cal S}_1$, and $\varphi_0$.

\subsection{Fitting the doped data}

\begin{figure}[t!]
\centering
\includegraphics[width=180mm]{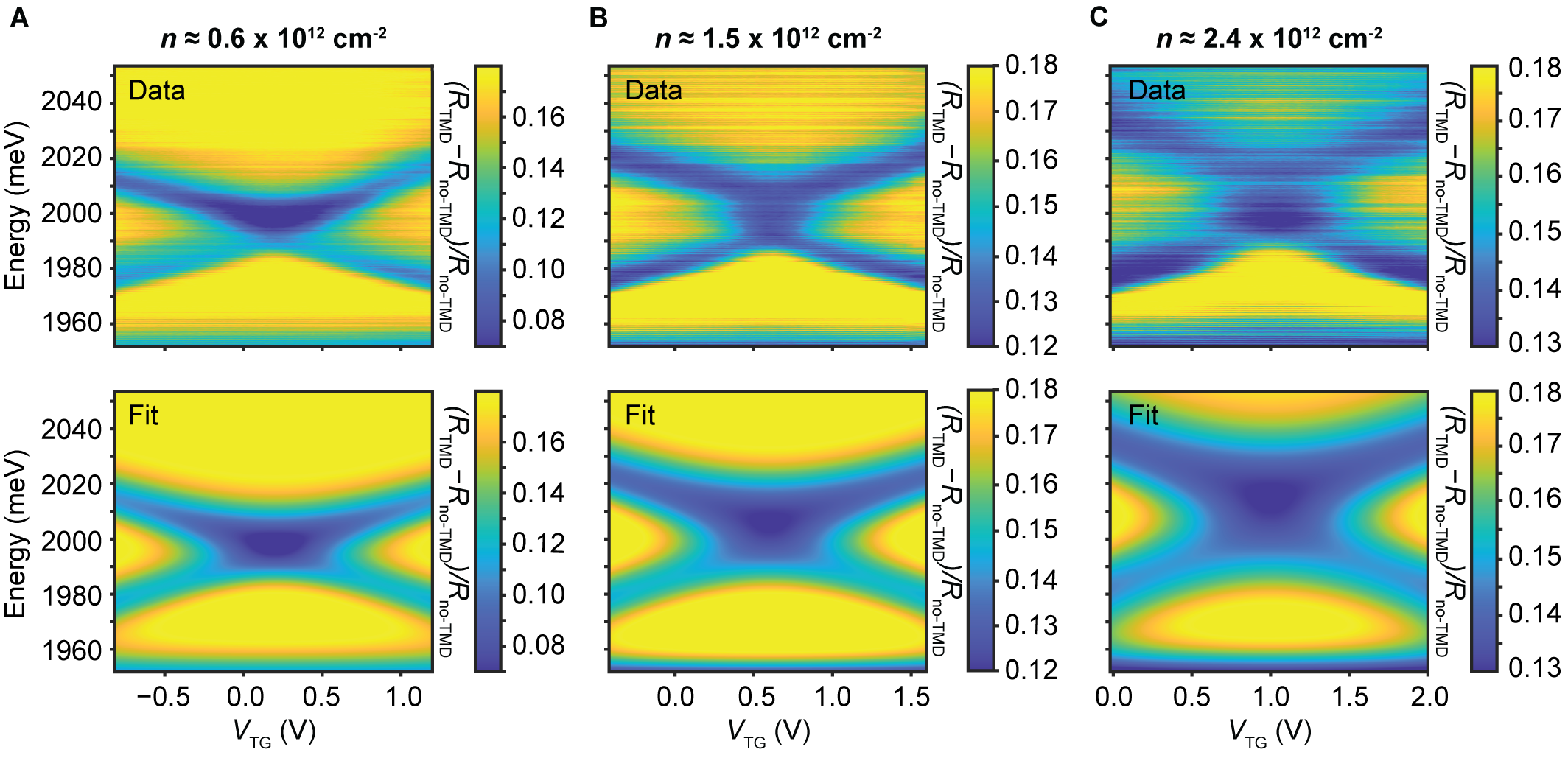} 
\caption{Data processing when the sample is doped.
\textbf{A}, \textbf{B}, \textbf{C} depict a back-to-back comparison of the measured (top panels) and fitted (bottom panels) electric-field sweeps at three representative dopings. This comparison demonstrates that the few-parameter fit is in reasonable quantitative agreement with the data, capturing both the linear Stark effect and the stochastic anti-crossing.
}
\label{fig::doped_fit}
\end{figure}

As discussed in Sec.~\ref{sec:finite W0}, we analyze interlayer excitons using full 2D reflectance maps and model IXs via Eqs.~\eqref{eqn:chi} and~\eqref{eqn:I_exprs}.
To minimize the number of fitting parameters while capturing both the linear Stark effect and stochastic hybridization, we represent these excitons with six parameters: ${\cal W}_0$, $\sigma$, $d_{\rm T} = d_{\rm B} = d$, $\gamma_{\rm T} = \gamma_{\rm B}$, and both the bare interlayer exciton energy $\omega_0$ and linear Stark shift, as encoded in $d_z$, are allowed to depend on the electron density. Figure~\ref{fig::doped_fit} shows that this few-parameter fit reasonably well captures the measured signal. While not as perfect as in Fig.~\ref{fig::intrinsic_fit}, this fit provides robust data processing (see below) by using significantly fewer parameters and still captures the essential physics.
This fitting procedure is then used to analyze the experiment (see Fig.~2 of the main text).

Let us comment on the physical content of each of the six fitting parameters associated with the interlayer excitons:
\begin{itemize}
    \item The mean coupling ${\cal W}_0$ encodes the intensity asymmetry between the upper and lower exciton branches.
    \item The stochastic variance ${\sigma}$ describes the stochastic anti-crossing, which is most evident at $E_z = 0$.
    \item The rate $\gamma$ corresponds to the excitonic linewidth for some appreciable $E_z \neq 0$.
    \item The transition dipole moment $d$ encodes the exciton-photon coupling and IX oscillator strength.
    \item $d_z$ is the effective IX out-of-plane dipole moment.
    \item $\omega_0$ is the bare IX energy at $E_z = 0$.
\end{itemize}

\subsection{Error bar analysis}
In estimating error bars for the fitted mean coupling ${\cal W}_0$ and stochastic variance $\sigma$, we consider both experimental error and fitting error.

The experimental error arises primarily from two factors: i) the inherent variability in individual measured spectra, as each spectrum represents an average of several measurements, leading to a variance in the measured signal, and ii) the signal-to-noise ratio, which we assess by smoothing the spectrum using a Savitzky-Golay filter and extracting the variance relative to the smoothed spectrum. However, we find that the experimental error is negligible compared to the fitting error. A complication we face with our fits is the worsening signal-to-noise ratio as the temperature increases.

\begin{figure}[t!]
\centering
\includegraphics[width=180mm]{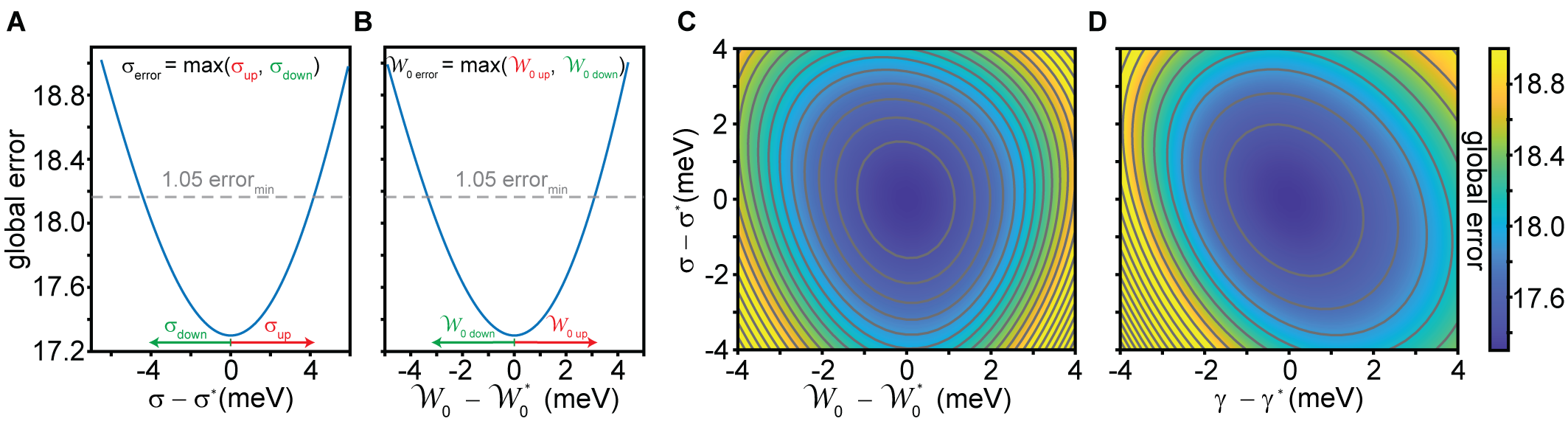} 
\caption{Error bar analysis of the stochastic variance $\sigma$ (\textbf{A}) and the mean coupling ${\cal W}_0$ (\textbf{B}). 
The error bars are estimated as follows: 
we scan $\sigma$ (or ${\cal W}_0$) near its optimized value $\sigma^*$ (${\cal W}^*_0$), while the rest of the parameters are refitted, and plot the resulting global least-squares error function. To determine the error bar, we use an error tolerance of 5\% above the global error minimum (indicated by dashed lines in \textbf{A} and \textbf{B}). \textbf{C} and \textbf{D} depict the global error function in the $({\cal W}_0,\sigma)$-map (\textbf{C}) and $(\gamma_0,\sigma)$-map (\textbf{D}), respectively, illustrating that the global error is reasonably isotropic in the vicinity of the optimized parameters.
}
\label{fig::error_bar}
\end{figure}

When using standard statistical tools, such as those based on confidence intervals ({\it lsqcurvefit} function in Matlab), we often obtain fitting errors that are unreasonably small. This situation arises because our few-parameter model is somewhat constrained, leading to an incidence of underfitting in the analysis of 2D reflectance maps~\cite{hastie2009elements,james2013introduction}. To estimate the fitting error bars for parameters like the stochastic variance $\sigma$, we then proceed as follows: we sweep $\sigma$ around the optimal value $\sigma^*$ while re-fitting the remaining parameters and evaluating the global least-squares error:
\begin{align}
\mathcal{F}[\bm{\theta}] = \sum_{i,j} \left| \mathcal{S}(\omega_i,V_{{\rm TG},j}) - \mathcal{S}_{\rm fit}[\bm{\theta}](\omega_i,V_{{\rm TG},j}) \right|^2,
\end{align}
where $\bm{\theta}$ represents the vector of fitting parameters (in our case, these are three parameters associated with the $A$-exciton and six with the interlayer excitons). Figure~\ref{fig::error_bar}A,B shows such scans for the stochastic variance $\sigma$ (A) and the mean coupling ${\cal W}_0$ (B), where, as expected, the global error displays a minimum at the optimized values $\sigma^*$ and ${\cal W}_0^*$, respectively. We define the error tolerance to be 5\% above the global error minimum, leading to error bars that significantly better represent the measured data (Fig.~2 of the main text).

To provide more insight in how well the fitting model represents the measured signal, we perform similar but now 2D scans of the global error as in Fig.~\ref{fig::error_bar}C,D. We find reasonably isotropic error contours indicative of i) the robustness of parameter estimates, ii) the consistency of our fitting, and iii) the independence of model parameters~\cite{hastie2009elements,james2013introduction}. These contours, thus, suggest that the model is well-suited to capturing the underlying data distribution and that the fitted parameters accurately represent the relationships between the model and the observed data.

\section{Theory for the stochastic anti-crossing}
\label{sec_SI_main_theory}

In the main text, we argued that the experimental conditions (low temperatures, strong Coulomb interactions $r_s \simeq 10-20$, small interlayer separation $lk_F \ll 1$, and absence of electron tunneling) strongly suggest the presence of interlayer electron coherence~\cite{zheng1997exchange,zhu2024interaction,stern2000dissipationless}. Here, we further discuss how such coherence can lead to the robust emergence of the stochastic anti-crossing.

In TMDs, the valley degree of freedom allows for interlayer electron coherence with intravalley character, $\Delta_{\rm intra, K}(\bm k) = \langle \hat{e}^\dagger_{\rm T,K}(\bm k ) \hat{e}_{\rm B,K}(\bm k) \rangle \neq 0$, and/or intervalley character, $\Delta_{\rm inter}(\bm k) = \langle \hat{e}^\dagger_{\rm T,K}(\bm k ) \hat{e}_{\rm B,K'}(\bm k) \rangle \neq 0$ (see Fig.~4A of the main text). 
Here, $\hat{e}^\dagger_{\rm T,K}(\bm k)$ is the electron creation operator in the top $K$-valley, with $\bm k$ being the wave vector relative to the band bottom.
For our discussion, we assume the system is spin-polarized but valley-depolarized, a situation expected in our finite magnetic-field measurements (see Fig.~3 of the main text). This assumption simplifies the analysis by omitting the electron spin, although including it is straightforward.
The close energies of these two order parameters make it likely that both are present and interplay within the system, as further discussed below.

\subsection{ Intravalley interlayer exciton hybridization and crude estimates}
\label{subsec:estimates}

Interlayer excitons serve as an optical probe of doped electrons; for this reason, we distinguish between the electron system with its microscopic interactions and the excitonic probe. Assuming the presence of interlayer electron coherence, we show here how this order parameter leads to interlayer exciton hybridization, consistent with our experimental findings, and discuss its stochastic behavior.

The absence of electron tunneling in MoS$_2$-homobilayers~\cite{pisoni2019absence} indicates that the many-body electron Hamiltonian (approximately, see Sec.~\ref{sec_SI_FS}) commutes with the particle number operators $\hat{N}_{\rm T}$ and $\hat{N}_{\rm B}$ in each layer.
Interlayer electron coherence corresponds to the spontaneous breaking of layer U(1) symmetry (in Sec.~\ref{sec_SI_FS}, we discuss that this U(1) symmetry is weakly broken down to $\mathds{Z}_2$), where the symmetry-broken state is not an eigenstate of $\hat{N}_{\rm T}-\hat{N}_{\rm B}$. This order parameter can be thought of as pseudospin ferromagnetism with the layer pseudospin pointing in-plane (see, for instance, Ref.~\cite{zheng1997exchange}).

The key idea of the hybridization mechanism we propose in Fig.~4B of the main text is that the presence of the intravalley coherence mediates strong electron tunneling-like processes. 
To illustrate this, we write the density-density interlayer Coulomb interaction between conduction-band electrons as (scattered electrons remain in the same valley and layer):
 \begin{align}
     \hat{H}_l =  \frac{1}{{\cal A}} \sum_{\bm k,\bm k',\bm q} \sum_{\alpha,\beta = {\rm K, K'} } {V}_l(q) \hat{e}^\dagger_{\rm T,\alpha}(\bm k + \bm q)\hat{e}_{\rm T,\alpha}(\bm k)
     \hat{e}^\dagger_{\rm B,\beta}(\bm k')\hat{e}_{\rm B,\beta}(\bm k' + \bm q),\label{eqn:H_l}
 \end{align}
where $V_l(q)$ is the interlayer Coulomb potential and ${\cal A}$ is the area of the sample.  
For estimates, we use $V_l(q) = 2\pi e^2 e^{- q l}/(\varepsilon q)$ with $\varepsilon \approx 3.76$ -- this form might slightly overestimate the strength of interlayer interactions as the TMD permittivity is larger than that of hBN~\cite{pisoni2019absence}. 
The presence of a nonzero order parameter, $\Delta_{\rm intra, K}(\bm k) = \langle \hat{e}^\dagger_{\rm T,K}(\bm k ) \hat{e}_{\rm B,K}(\bm k) \rangle \neq 0$, results in an effective electron tunneling-like term (we write only the processes in the $K$-valley):
\begin{align}
     -\sum_{\bm k} t_{\bm k} 
     \hat{e}^\dagger_{\rm B,K}(\bm k)\hat{e}_{\rm T,K}(\bm k) + \text{h.c.},
 \end{align}
where the effective tunneling rate is set by the order parameter:
\begin{align}
    t_{\bm k} = \frac{1}{{\cal A}} \sum_{\bm q} V_l(\bm q) \Delta_{\rm intra, K}({\bm k + \bm q}).\label{eqn:tun_eff}
\end{align}
Let us note that a small but finite electron tunneling, which explicitly breaks the U(1) layer symmetry, would imprint the phase on the order parameter, much like a small magnetic field in a ferromagnet polarizes spins along its direction.
In spin systems with strong ferromagnetic correlations, a small magnetic field induces significant spin polarization. Analogously, in our case, even weak interlayer electron tunneling is expected to enhance tunneling conductance. This effect, a definitive signature of interlayer electron coherence, has been experimentally established in conventional quantum Hall bilayers~\cite{spielman2000resonantly,lin2022emergence,wen1993tunneling,stern2000dissipationless,stern2001theory,fogler2001josephson}.

The layer separation in our system is only a few angstroms, making the effective electron tunneling-like processes in Eq.~\eqref{eqn:tun_eff} strong.  Assuming perfect Hartree-Fock correlations with $\Delta_{\rm intra, K}(k) \simeq n_F(k)$ ($n_F$ is the Fermi-Dirac distribution function), we estimate $t_{\bm k = \bm 0} \simeq e^2 \sqrt{\pi n}/\varepsilon \simeq 96\,$meV and $t_{k = 1/a_X} \simeq 36\,$meV for four electron bands (corresponding to the spin-polarized case as in Fig.~3 of the main text), $n = 2\times 10^{12}\,$cm$^{-2}$, $a_X = 3\,$nm, and $T = 0\,$K -- see also Sec.~\ref{subsec:scHF}, where we detail our self-consistent Hartree-Fock analysis.

The significance of such tunneling-like electron processes is that they give rise to the hybridization of, for example, IX$_{\rm T}$- and $A_{\rm B}$-excitons -- see Fig.~4B (left) of the main text.
This coupling could be estimated as (we note that $l \ll a_X$):
\begin{align}
    t_{\rm IX_{\rm T} \leftrightarrow A_{\rm B}} \simeq \int \frac{d^2 \bm k}{(2\pi)^2} t_{\bm k} \Psi^*_A(\bm k)\Psi_X(\bm k) \simeq 85\,\text{meV for  $n = 2\times 10^{12}\,$cm$^{-2}$},
    \label{eqn: t_int}
\end{align}
where we substituted for the exciton wave-functions
$\Psi_A(\bm q) \approx \Psi_X(\bm q) \approx 2 \sqrt{2\pi} a_X/((qa_X)^2 + 1)^{3/2}$. 
Given our assumptions, the value in Eq.~\eqref{eqn: t_int} is likely an overestimate, but it nevertheless underscores the importance of the processes we propose.
At the same time, the $A_{\rm B}$-exciton couples to the IX$_{\rm B}$-state via the two-step process shown in Fig.~4B (middle and right panels) of the main text, with experimental evidence supporting this effect~\cite{sponfeldner2022capacitively} and a coupling strength of about $\sim 4$ meV. All three processes in Fig.~4B of the main text combined lead to the intravalley interlayer exciton hybridization (we again write only the $K$-valley terms):
\begin{align}
\hat{H}_{\rm intravalley} =  \delta {\cal W}[\Delta_{\rm intra, K}] \hat{ X}^\dagger_{\rm B,K}\hat{ X}_{\rm T,K} + \text{h.c.}
\label{eqn:H_eff_IX_v0}
\end{align}
Using second-order perturbation theory, the coupling $\delta {\cal W}[\Delta_{\rm intra, K}]$ between the interlayer excitons can then be estimated as $85\,\text{meV}\times 4\,\text{meV}/70\,\text{meV} \simeq 5\,$meV for $n = 2\times 10^{12}\,$cm$^{-2}$ (here, $70\,\text{meV}$ is the energy difference between IX- and $A$-excitons). 
While this estimate is rather crude -- as we (i) assumed perfect Hartree-Fock correlations, (ii) considered only the lowest energy intermediate exciton states (we note that the 2$s$ $A$-exciton, though having a small oscillator strength, is energetically closer to the interlayer excitons, see also Sec.~\ref{sec_SI_FS}), and (iii) used perturbation theory to relate the electronic order parameter to interlayer exciton hybridization -- we find that the estimated value is comparable to the measured ones (Fig.~2E of the main text), suggesting that the proposed hybridization mechanism is realistic.

\subsection{Additional symmetry considerations for MoS$_2$-homobilayers}
\label{sec_main_symm}

In contrast to conventional semiconductors with a single electron valley (assuming the electron system is spin-polarized), MoS$_2$-homobilayers feature not just a single but two distinct $K$- and $K'$-valleys (when the electron spin can be disregarded, these valleys are approximately degenerate), which bring in an additional spin-$1/2$-like degree of freedom. 
The effective microscopic Hamiltonian, consisting of the electron kinetic energy (with approximately parabolic dispersion) in each of the valleys and Coulomb interactions, now commutes with the particle number operators  $\hat{N}^\alpha_{\rm T/B} = {\cal A}^{-1}\sum_{\bm k} \hat{e}^\dagger_{\rm T/B,\alpha}(\bm k)\hat{e}_{\rm T/B,\alpha}(\bm k)$ in each layer and each valley $\alpha \in \{ {\rm K,K'} \}$. 
The valley degree of freedom suggests the introduction of the order parameter as:
\begin{align}
    \hat{e}^\dagger_{\rm T,\alpha}(\bm k)  \hat{e}_{\rm B,\beta}(\bm k)  \to \hat{\Delta}_0(\bm k) \tau^0_{\alpha\beta} + [\hat{\Delta}_x(\bm k) \tau^x_{\alpha\beta} + \hat{\Delta}_y(\bm k) \tau^z_{\alpha\beta} + \hat{\Delta}_z(\bm k) \tau^z_{\alpha\beta}],
    \label{eqn:OP_full}
\end{align}
 where $\hat{\Delta}_a (\bm k) \equiv \frac{1}{2} \hat{e}^\dagger_{\rm T,\alpha}(\bm k) \tau^a_{\alpha \beta}  \hat{e}_{\rm B,\beta}(\bm k)$,
$a \in \{ 0,x,y,z\}$, and $\tau^a$ are the Pauli matrices in the valley-space. 
The $K$-valley electron coherence, which determines the $K$-valley interlayer exciton hybridization in Eq.~\eqref{eqn:H_eff_IX_v0}, see Sec.~\ref{subsec:estimates}, is then written as $\Delta_{\rm intra, K}(\bm k) = \langle \hat{\Delta}_0(\bm k) \rangle + \langle \hat{\Delta}_z(\bm k) \rangle $. 
We note that the component $\hat{\Delta}_0$ transforms trivially (as a scalar) under the global SU(2) valley rotations, while the vector components $(\hat{\Delta}_{x},\hat{\Delta}_{y},\hat{\Delta}_{z})$ transform as an SU(2) triplet.
Therefore, if the electronic ground state develops $\langle \hat{\Delta}_0 \rangle \neq 0$ only, it breaks the original U(1)$\times$SU(2) symmetry -- here, the U(1) part is associated with the operator $\hat{N}^{\rm tot}_{\rm T} - \hat{N}^{\rm tot}_{\rm B} = \hat{N}^{\rm K}_{\rm T} + \hat{N}^{\rm K'}_{\rm T} - \hat{N}^{\rm K}_{\rm B} - \hat{N}^{\rm K'}_{\rm B}$, while the SU(2) part is associated with valley pseudospin rotations -- down to SU(2). On the other hand, the development of a nonzero vector component breaks both the U(1) and SU(2) parts of this U(1)$\times$SU(2) symmetry.

This effective microscopic description, where the valley index is analogous to a spin index, implies that the system can be thought of as effectively translationally invariant (noting that in the sample, intervalley correlations actually carry the momentum $\bm K - \bm K'$). 
For this reason,  in Eq.~\eqref{eqn:OP_full}, we consider both electron operators to have the same momentum.
In other words, we assume that the ground state can give a nonzero expectation value $\langle \hat{e}^\dagger_{\rm T,\alpha}(\bm k)  \hat{e}_{\rm B,\beta}(\bm k') \rangle \neq 0$ only if $\bm k = \bm k'$.
With this in mind, we write the reduced~\cite{girvin2019modern} (corresponding to $\bm k = \bm k'$ in Eq.~\eqref{eqn:H_l}) interlayer Coulomb interaction~\eqref{eqn:H_l} as:
\begin{align}
    \hat{H}_l \to - \frac{2}{ {\cal A}} \sum_{\bm k,\bm k'} V_l(|\bm k - \bm k'|) [ \hat{\Delta}_0^\dagger(\bm k) \hat{\Delta}_0(\bm k') + \hat{\Delta}_x^\dagger(\bm k) \hat{\Delta}_x(\bm k') + \hat{\Delta}_y^\dagger(\bm k) \hat{\Delta}_y(\bm k')+ \hat{\Delta}_z^\dagger(\bm k) \hat{\Delta}_z(\bm k') ],
    \label{eqn:H_l_red}
\end{align}
and this form is clearly U(1)$\times$SU(2) symmetric. 
This form also indicates that the intravalley and intervalley electron coherences stand on equal footing in our system. 

At the same time, from the perspective of intravalley interlayer excitons, the coupling in Eq.~\eqref{eqn:H_eff_IX_v0} requires the explicit presence of a nonzero order parameter. Indeed, within the perturbation theory discussed above, we get
\begin{align}
    \delta \hat{\cal W}_{\rm intra, K} \propto t_h   \int \frac{d^2 \bm k}{(2\pi)^2}\int \frac{d^2 \bm q}{(2\pi)^2} \Psi^*_A(\bm k)\Psi_X(\bm k)  V_l(\bm q) [\hat{\Delta}_0({\bm k + \bm q}) + \hat{\Delta}_z({\bm k + \bm q})],\label{eqn:W_intra_OP}
\end{align}
where $t_h$ is the hole tunneling rate, on the order of a few tens of meV in MoS$_2$-homobilayers. (From the symmetry perspective, holes are allowed to tunnel as the valence bands have the same AQNs, see also  the right panel of Fig.~4B in the main text.)
The expression~\eqref{eqn:W_intra_OP} indicates that $\delta \hat{\cal W}_{\rm intra, K}$ is sensitive to both the order parameter amplitude and phase, thereby inheriting spatial inhomogeneities due to statistical fluctuations of the order parameter phase. 
Consequently, the model in Eq.~\eqref{eqn:H_eff_IX_v0} should be extended to account for spatial dependence.
In the experiment, this coupling is spatially averaged over the optical spot size, which we expect to be much larger than the phase coherence length (see also Sec.~\ref{sec_SI_spot_size}). 
As a result, the spatial average $\langle\delta {\cal W}[\Delta_{\rm intra, K}]\rangle \approx 0$ vanishes, but the variance $\sigma$ can be significant, manifesting as the stochastic anti-crossing. This explains the experimental observations, as elaborated in the main text.

We comment that MoS$_2$-homobilayers possess ${\cal C}_3$-rotationally rotational symmetry, assigning opposite AQNs to interlayer excitons within the same valley. 
As a result, there is no Hamiltonian term that directly couples such two excitons (disorder could potentially play a role, but as discussed in the main text, it is not expected to be dominant).
These excitons can hybridize when the system breaks this symmetry, specifically through intravalley interlayer electron coherence. In other words, the coupling in Eq.~\eqref{eqn:H_eff_IX_v0} serves as a probe of this order parameter.
We also note that because these two excitons have opposite AQNs, their hybridization cannot explain the small intensity asymmetry observed between the lower and upper exciton branches -- see Fig.~\ref{fig::3dopingSweeps} and Sec.~\ref{sec:finite W0}. We address this question further in the following section.

\subsection{Self-consistent Hartree-Fock analysis}
\label{subsec:scHF}

We conclude this section by presenting our self-consistent Hartree-Fock analysis, which reasonably captures the observations in Fig.~2E,F of the main text.
These calculations extend the single-band analysis in Refs.~\cite{zheng1997exchange,zhu2024interaction} to the case of two $K$- and $K'$-valleys relevant for TMDs. Specifically, we consider the following microscopic Hamiltonian (neglecting electron spin and, as such, the small spin-orbit coupling):
\begin{align}
    \hat{H} & = \sum_{\bm k} \sum_{l,v} \frac{k^2}{2 m^*} \hat{e}^\dagger_{lv}(\bm k)\hat{e}_{lv}(\bm k)  + \frac{1}{2{\cal A}} \sum_{\bm k,\bm k',\bm q} \sum_{l }  \sum_{vv'} {V}(q) \hat{e}^\dagger_{lv}(\bm k + \bm q)
     \hat{e}^\dagger_{lv'}(\bm k' - \bm q)\hat{e}_{lv'}(\bm k')\hat{e}_{lv}(\bm k) \notag \\
     & \qquad\qquad\qquad\qquad\qquad\qquad\qquad\qquad\qquad\qquad\quad
     + \frac{1}{{\cal A}} \sum_{\bm k,\bm k',\bm q}  \sum_{vv'} {V}_l(q) \hat{e}^\dagger_{{\rm T}v}(\bm k + \bm q)
     \hat{e}^\dagger_{{\rm B}v'}(\bm k' - \bm q)\hat{e}_{{\rm B}v'}(\bm k')\hat{e}_{{\rm T}v}(\bm k), \label{eqn:H_full}
\end{align}
where $l\in\{ {\rm T,B} \}$ is the layer index, $v\in\{ {K,K'} \}$ is the valley index, and $m^*$ is the effective electron mass.

Within the Hartree-Fock approximation, and in the strongly-interacting regime relevant to the experiment ($1 \ll r_s$), the ground-state wave function with $\langle\hat{\Delta}_0\rangle \neq 0$ is given by (analogous to the $\ket{S_2}$-state in the single-band case~\cite{zheng1997exchange,zhu2024interaction}):
\begin{align}
    \ket{\psi_0} =  \prod_{k\leq k_F} \frac{1}{2}(\hat{e}^\dagger_{\rm T,K}(\bm k) + \hat{e}^\dagger_{\rm B,K}(\bm k))(\hat{e}^\dagger_{\rm T,K'}(\bm k) + \hat{e}^\dagger_{\rm B,K'}(\bm k))\ket{0},\label{eqn:wf_psi_0}
\end{align}
where $k_F = \sqrt{2\pi n}$. For future reference, we define $T_F = k_F^2/(2m^*) = \pi n/m^* = (2/r^2_s)\text{Ry}^*$, where $r_s =1/(a^* \sqrt{\pi n})$, $\text{Ry}^* \equiv  e^2/(2a^*\varepsilon) = 1/(2m^* (a^*)^2)$ is the Rydberg energy, and $a^* \equiv \varepsilon/(m^* e^2)$ is the Bohr radius (estimated to be $a^*\simeq 0.3\,$nm and $\text{Ry}^*\simeq 530\,$meV).

We propose that this many-body ground state wave funcion~\eqref{eqn:wf_psi_0} can qualitatively explain our low-temperature measurements shown in Fig.~2E of the main text. To support this, we evaluate the expression in Eq.~\eqref{eqn: t_int}, which captures the effective hybridization strength between the IX$_{\rm T}$- and $A_{\rm B}$-excitons:
\begin{align}
    t_{\rm IX_{\rm T} \leftrightarrow A_{\rm B}} & \simeq \int \frac{d^2 \bm k}{(2\pi)^2} \frac{8 \pi a_X^2}{(1 + (k a_X)^2)^3}  \int \frac{d^2 \bm q}{(2\pi)^2} V_l(\bm q) \langle \hat{e}^\dagger_{\rm T,K}(\bm k + \bm q)\hat{e}_{\rm B,K}(\bm k + \bm q) \rangle\\
    & \simeq 
    \int \frac{d^2 \bm k}{(2\pi)^2} \frac{8 \pi a_X^2}{(1 + (k a_X)^2)^3} \frac{2\pi e^2}{k \varepsilon}  \int \frac{d^2 \bm q}{(2\pi)^2} \langle \hat{e}^\dagger_{\rm T,K}(\bm q)\hat{e}_{\rm B,K}(\bm q)\rangle 
    \\
    & =\frac{3\pi^2e^2 a_X}{2\varepsilon} \int \frac{d^2 \bm q}{(2\pi)^2} \langle \hat{e}^\dagger_{\rm T,K}(\bm q)\hat{e}_{\rm B,K}(\bm q)\rangle  = \frac{3 \pi^2 e^2a_X}{8\varepsilon} n , \label{eqn:T=0_tb}
\end{align}
where in the second identity, we have used $k_F \ll a_X^{-1} \ll l^{-1}$, a condition that further implies that the rate $ t_{\rm IX_{\rm T} \leftrightarrow A_{\rm B}} \propto \Delta_0(\bm r =0)$ is approximately set by the local value of the order parameter $ \Delta_0(\bm r)$. 
Our mean-field analysis neglects spatial inhomogeneities in the order parameter due to phase fluctuations. Thus, using Eq.~\eqref{eqn:T=0_tb} and perturbation theory for interlayer exciton hybridization described above, our Hartree-Fock analysis estimates the stochastic variance:
\begin{align}
    \sigma_{\rm mf} = C \int \frac{d^2 \bm q}{(2\pi)^2} \langle \hat{e}^\dagger_{\rm T,K}(\bm q)\hat{e}_{\rm B,K}(\bm q)\rangle.
\end{align}
where in our approach, $C$ is a phenomenological parameter that may depend on temperature $T$. 
Notably, at low temperatures, $\sigma_{\rm mf}$ is approximately proportional to the electron density $n$, Eq.~\eqref{eqn:T=0_tb}, consistent with our measurements in Fig.~2E of the main text (dashed line), which were used to determine $C$. 
Let us comment that while the Hartree-Fock approximation might underestimate the role of low-momenta fluctuations, it should reasonably well capture local properties, particularly the stochastic variance $\sigma$.

\begin{figure}[t!]
\centering
\includegraphics[width=180mm]{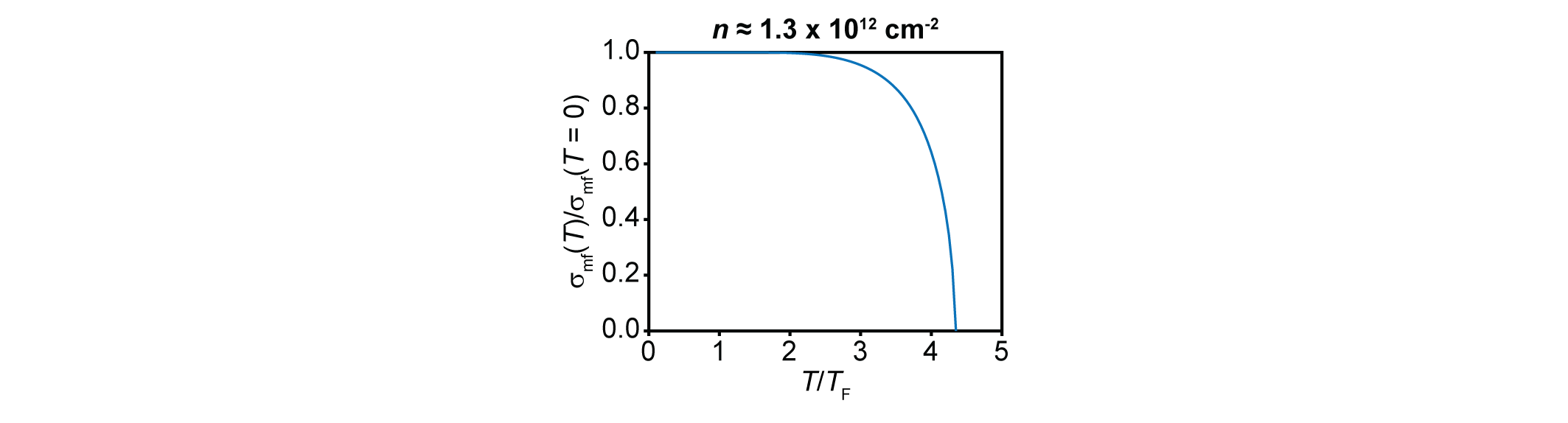} 
\caption{
Temperature-dependent, self-consistent Hartree-Fock simulation at $n \approx 1.3 \times 10^{12}\,\text{cm}^{-2}$.
}
\label{fig::theory_exp_2b}
\end{figure}

To explore finite-temperature effects within the same mean-field approximation, where the many-body density matrix is assumed to be Gaussian, we introduce the covariance matrix:
\begin{align}
    \Gamma_{vv'}^{ll'}(\bm k) \equiv \langle \hat{e}^\dagger_{lv}(\bm k)\hat{e}_{l'v'}(\bm k) \rangle.
\end{align}
For the state in Eq.~\eqref{eqn:wf_psi_0}, the covariance matrix reads:
\begin{align}
    \Gamma_{\bm k}\Big|_{k\leq k_F} = \frac{1}{2}\begin{bmatrix}
        \mathbb{1}_{2\times 2} & \mathbb{1}_{2\times 2}\\
        \mathbb{1}_{2\times 2} & \mathbb{1}_{2\times 2}
    \end{bmatrix}. 
\end{align}
We note that this matrix satisfies the purity condition $ \Gamma_{\bm k}^2 =  \Gamma_{\bm k}$, consistent with the fact that the density matrix at $T = 0$ is pure.
The energy expectation value is then understood as:
\begin{align}
    E[\Gamma] = \sum_{\bm k} \frac{k^2}{2m^*} \text{tr}(\Gamma(\bm k))
    - \frac{1}{2{\cal A}}\sum_{\bm k,\bm k'}\sum_l \sum_{vv'} V(\bm k-\bm k')\Gamma_{vv'}^{ll}(\bm k)\Gamma_{v'v}^{ll}(\bm k') - \frac{1}{{\cal A}}\sum_{\bm k,\bm k'}\sum_{vv'} V_l(\bm k-\bm k')\Gamma_{vv'}^{\rm TB}(\bm k)\Gamma_{v'v}^{\rm BT}(\bm k')
    ,\label{eqn:E_Gamma_2b}
\end{align}
where the first term represents the kinetic energy, and the last two terms correspond to the exchange energies associated with intra- and interlayer electron interactions, respectively. 
The electrostatic energy is not included as it vanishes for the balanced symmetric case of equal densities considered here~\cite{zheng1997exchange,zhu2024interaction,PhysRevB.61.13882}.
Specifically, we assume that (i) all four conduction bands (two valleys and two layers) are equally populated (consistent with the magnetic field measurements discussed in the main text) and (ii) there is no intralayer intervalley coherence. Under these assumptions, and using similar notation as above, the covariance matrix takes the form:
\begin{align}
    \Gamma(\bm k) = n_{\bm k} \mathbb{1}_{4\times 4} + \begin{bmatrix}
        \mathbb{0}_{2\times 2} & \Delta_0(\bm k) \tau^0 + \Delta_x(\bm k) \tau^x + \Delta_y(\bm k) \tau^y + \Delta_z(\bm k) \tau^z \\
        \Delta_0^*(\bm k) \tau^0 + \Delta_x^*(\bm k) \tau^x + \Delta_y^*(\bm k) \tau^y + \Delta_z^*(\bm k) \tau^z  & \mathbb{0}_{2\times 2}
    \end{bmatrix},\label{eqn:Gamma_OP_2b}
\end{align}
where $\Delta_a(\bm k) = \frac{1}{2}\text{tr}( \Gamma^{\rm TB}(\bm k)\tau^a)$. For such states, the energy expectation value is further given by, cf. Eq.~\eqref{eqn:H_l_red}:
\begin{align}
    E[\Gamma] & =  \sum_{\bm k} \frac{2 k^2n_{\bm k}}{m^*}  
    - \frac{2}{{\cal A}}\sum_{\bm k,\bm k'} V(\bm k-\bm k')n_{\bm k}n_{\bm k'}\notag\\
    &
    - \frac{2}{{\cal A}}\sum_{\bm k,\bm k'} V_l(\bm k-\bm k')[\Delta_0^*(\bm k')\Delta_0(\bm k) + \Delta_x^*(\bm k')\Delta_x(\bm k) + \Delta_y^*(\bm k')\Delta_y(\bm k) + \Delta_z^*(\bm k')\Delta_z(\bm k)].\label{eqn:E_Gamma_2b_v2}
\end{align}
This expression reflects that the Hamiltonian in Eq.~\eqref{eqn:H_full} is U(1)$\times$SU(2) symmetric, as discussed above.

The mean-field Hamiltonian $\hat{H}_{\rm MF} = \sum_{\bm k}\hat{\Psi}_{\bm k}^\dagger h({\bm k}) \hat{\Psi}_{\bm k}$, where $\hat{\Psi}_{\bm k}\equiv (\hat{e}_{\rm T,K}(\bm k),\hat{e}_{\rm T,K'}(\bm k),\hat{e}_{\rm B,K}(\bm k),\hat{e}_{\rm B,K'}(\bm k))^{\rm T}$, is derived from Eq.~\eqref{eqn:E_Gamma_2b} by evaluating the variational derivative $h_{vv'}^{ll'}(\bm k) = \delta E[\Gamma]/\delta \Gamma_{vv'}^{ll'}(\bm k)$ and can be expressed as:
\begin{align}
    h_{vv'}^{ll'}(\bm k) = \frac{k^2}{2m^*}\delta_{ll'}\delta_{vv'} - \frac{\delta_{ll'}}{\cal A}\sum_{\bm k'}V(\bm k - \bm k')\Gamma^{ll}_{v'v} (\bm k')
    - \frac{(1 -\delta_{ll'})}{\cal A}\sum_{\bm k'}V_l(\bm k - \bm k')\Gamma^{l'l}_{v'v} (\bm k').
\end{align}
Without loss of generality, and as follows from Eq.~\eqref{eqn:E_Gamma_2b_v2}, we can consider states that can have $\Delta_0(\bm k) \neq 0$ only (we also fix $\Delta_0(\bm k)$ to be real so that the pseudospin points in-plane, along the $x$-axis), in which case the mean-field Hamiltonian is written as:
\begin{align}
    h(\bm k) =   \xi_{\bm k}\mathbb{1}_{4\times 4} - t_{\bm k}  \begin{bmatrix}
        \mathbb{0}_{2\times 2} & \mathbb{1}_{2\times 2}\\
        \mathbb{1}_{2\times 2} & \mathbb{0}_{2\times 2}
    \end{bmatrix},\label{eqn:h_fin}
\end{align}
where
\begin{align}
    \xi_{\bm k} = \frac{k^2}{2m^*}  - \frac{1}{\cal A}\sum_{\bm k'}V(\bm k - \bm k')n_{\bm k'},\qquad t_{\bm k} = \frac{1}{\cal A}\sum_{\bm k'}V(\bm k - \bm k')\Delta_0({\bm k'}). \label{eqn:sc_p1}
\end{align}
We write the eigenvectors of Eq.~\eqref{eqn:h_fin} as:
\begin{align}
    \begin{bmatrix}
        \hat{\psi}_{1,-}(\bm k)\\
        \hat{\psi}_{2,-}(\bm k)\\
        \hat{\psi}_{1,+}(\bm k)\\
        \hat{\psi}_{2,+}(\bm k)
    \end{bmatrix}
    = \frac{1}{\sqrt{2}} \begin{bmatrix*}[r]
        1 & 0 & 1 & 0\\
        0 & 1 & 0 & 1\\
        1 & 0 & -1 & 0\\
        0 & 1 & 0 & -1
    \end{bmatrix*}
    \begin{bmatrix}
        \hat{e}_{\rm T,K}(\bm k)\\
        \hat{e}_{\rm T,K'}(\bm k)\\
        \hat{e}_{\rm B,K}(\bm k)\\
        \hat{e}_{\rm B,K'}(\bm k)
    \end{bmatrix},
\end{align}
and the corresponding eigenvalues are $\xi_{\bm k} - t_{\bm k}$, $\xi_{\bm k} - t_{\bm k}$, $\xi_{\bm k} + t_{\bm k}$, and $\xi_{\bm k} + t_{\bm k}$, respectively. Self-consistency then implies:
\begin{align}
    n_{\bm k} = \frac{1}{2}[n_{F}(\xi_{\bm k} - t_{\bm k}) + n_{F}(\xi_{\bm k} + t_{\bm k})], 
    \qquad
    \Delta_0({\bm k}) = \frac{1}{2}[n_{F}(\xi_{\bm k} - t_{\bm k}) - n_{F}(\xi_{\bm k} + t_{\bm k})],\label{eqn:sc_p2}
\end{align}
where $n_F(\varepsilon) = [1 + \exp((\varepsilon - \mu)/T)]^{-1}$ is the Fermi-Dirac distribution function, and the chemical potential $\mu$ is set by the total density $n = \displaystyle 2\int\frac{d^2\bm k}{(2\pi)^2}[n_F(\xi_{\bm k} - t_{\bm k}) + n_F(\xi_{\bm k} + t_{\bm k}) ) ]$.

We solve Eqs.~\eqref{eqn:sc_p1} and~\eqref{eqn:sc_p2} numerically, with the results shown in Fig.~\ref{fig::theory_exp_2b}(b) using experimental parameters at $n = 1.3\times 10^{12}\,\text{cm}^{-2}$, i.e, as in Fig.~2F of the main text. 
The Hartree-Fock approximation predicts a critical temperature of $T_c \simeq 4.35 T_F \simeq 225\,\text{K}$; however, this value should be considered an upper bound, as this method tends to underestimate the role of low-momenta fluctuations. By rescaling the temperature to match the experimental critical temperature $T_c = 75\,\text{K}$ and using the experimentally determined value of $C$ (assumed to be temperature independent), we achieve reasonable agreement between the theory and the data, as shown in Fig.~2F of the main text. 
This analysis suggests that the disappearance of the stochastic anti-crossing can be understood as the order parameter amplitude is suppressed with increasing $T$ until it eventually melts.

\section{ Intervalley interlayer exciton hybridization,
weak symmetry breaking, and Fermi sea fluctuations}
\label{sec_SI_FS}

Our theory in Sec.~\ref{sec_SI_main_theory} provides an interpretation of essentially all the experimental features except the small asymmetry between the lower and upper exciton branches -- see, for instance, Fig.~\ref{fig::3dopingSweeps}. 
In Sec.~\ref{sec:finite W0}, we argued that this asymmetry is indicative of intervalley interlayer exciton hybridization (as opposed to intravalley exciton hybridization discussed in Sec.~\ref{sec_SI_main_theory}), which have the same AQNs.
In the experiment, this asymmetry is associated with the development of small mean value ${\cal W}_0 \neq 0$, which has large error bars -- see Fig.~2E,F of the main text.
Furthermore, in Sec.~\ref{sec_SI_spot_size} we further experimentally argue that this mean value ${\cal W}_0 \neq 0$ does not originate from order parameter phase coherence.

Because these excitons -- such as $X_{\rm B,K}$ and $X_{\rm T,K'}$ (depicted in red in Fig.~1B of the main text) -- have the same AQNs, there is no symmetry argument preventing their direct hybridization even in the absence of doped electrons.
However, such a direct coupling is expected to be weak because the involved processes require both exciton electron and hole layer and valley switching. This expectation aligns with the experiment in Fig.~2E, which indicates that $|{\cal W}_0|\lesssim 2\,$meV for $n = 0$. 
We note that the interlayer excitons have rather large linewidths on the order of 10$\,$meV, making it difficult to resolve a small possible hybridization.
It is plausible that doped carriers could amplify this coupling through simple processes that do not involve the exotic physics discussed in the preceding section. To illustrate this expectation, in this section, we provide one such mechanism based on dynamical Fermi liquid fluctuations (we also mentioned in the main text that polaronic dressing might be important for the intervalley scenario as well).
Let us remark that while the existence of such dynamical processes can explain the weak exciton intensity asymmetry, they cannot account for the static stochastic variance $\sigma$.

To understand how intervalley interlayer excitons could hybridize, we introduce processes termed `hole flip' and `electron flip', both corresponding to layer switching and scattering across the TMD Brillouin zone -- see Fig.~\ref{fig::agnostic_flip_flop}.
The hole flip, shown in Fig.~\ref{fig::agnostic_flip_flop} (left), can occur through the simultaneous scattering of the hole of the $X_{\rm T,K'}$-exciton from the bottom $K'$-valley to the top $K$-valley and a Fermi sea electron. 
In such scattering processes, the total momentum is conserved, and the spin and AQN of the Fermi sea electron remain unchanged, resulting in the two possibilities depicted in Fig.~\ref{fig::agnostic_flip_flop} (left).
Figure~\ref{fig::agnostic_flip_flop} (right) shows that similar Fermi sea scatterings can give rise to the electron flip, allowing us to write the following bare microscopic Hamiltonian:
\begin{gather}
    \hat{H}_{\rm int}  = \frac{V_a}{{\cal A}} \sum_{\bm k,
    \bm k', \bm q} 
    ( \hat{\cal F}^{(e)}_{\bm k + \bm q,\bm k} \hat{e}^\dagger_{\rm B,K,\uparrow}(\bm k' - \bm q)\hat{e}_{\rm T,K',\uparrow}(\bm k') -  \hat{\cal F}^{(h)}_{\bm k + \bm q,\bm k}\hat{h}^\dagger_{\rm T,K,\downarrow}(\bm k' - \bm q)\hat{h}_{\rm B,K',\downarrow}(\bm k')) + \text{h.c.},\label{eqn:H_a_p1}
    \\
    \hat{\cal F}^{(e)}_{\bm k_1,\bm k_2}  \equiv \sum_\sigma [\hat{e}^\dagger_{\rm B,K',\sigma}(\bm k_1 )\hat{e}_{\rm T,K,\sigma}(\bm k_2)  + \hat{e}^\dagger_{\rm T,K',\sigma}(\bm k_1)\hat{e}_{\rm B,K,\sigma}(\bm k_2) ],\label{eqn:H_a_p2}\\
    \hat{\cal F}^{(h)}_{\bm k_1,\bm k_2} \equiv \sum_\sigma [ \hat{e}^\dagger_{\rm T,K,\sigma}(\bm k_1)\hat{e}_{\rm B,K',\sigma}(\bm k_2)  + \hat{e}^\dagger_{\rm B,K,\sigma}(\bm k_1)\hat{e}_{\rm T,K',\sigma}(\bm k_2)],\label{eqn:H_a_p3}
\end{gather}
where the hole creation operator is understood as $\hat{h}^\dagger_{\rm T/B,\sigma}(\bm q)\equiv \hat{e}_{{\rm T/B},v,\bar{\sigma}}(-\bm q)$. 
The parameter $V_a$ is determined by the Coulomb potential at $|\bm K - \bm K'|$ ($2\pi e^2/\varepsilon|\bm K - \bm K'|$, see also Refs.~\cite{zak2012ferromagnetic,miserev2019exchange}) and, since both scattered particles switch layers, by the corresponding wave function overlaps. Given the small interlayer separation $l\simeq 0.6\,$nm and strong hole tunneling (on the order of tens of meV), 
these overlaps can be non-negligible. In our approach, $V_a$ is a phenomenological parameter that is further assumed to be momentum-independent.

In what follows, we demonstrate that Eq.~\eqref{eqn:H_a_p1} leads to the intervalley interlayer exciton hybridization of the form:
\begin{align}
    \hat{H}_{\rm intervalley} =  \hat{\cal W}_{\rm inter} \hat{ X}^\dagger_{\rm B,K}\hat{ X}_{\rm T,K'} + \text{h.c.}
\label{eqn:H_inter}
\end{align}
We evaluate the strength $\hat{\cal W}_{\rm inter}$ within second-order perturbation theory, cf. Eq.~\eqref{eqn:W_gen}, and show that its mean expectation value can be nonzero $\langle\hat{\cal W}_{\rm inter}\rangle \neq 0$ even if the system does not spontaneously break any of the symmetries mentioned in Sec.~\ref{sec_SI_main_theory}, thereby providing an interpretation of the weak intensity asymmetry in, for example, Fig.~\ref{fig::3dopingSweeps}, as well as optical size effects discussed in Sec.~\ref{sec_SI_spot_size}.

\subsection{Anomalous terms and the U(1) layer symmetry}

Before we proceed, let us mention that strictly speaking, the Hamiltonian in Eq.~\eqref{eqn:H_a_p1} contains terms such as $\sim \hat{e}^\dagger_{\rm T,K}\hat{e}^\dagger_{\rm T,K'}\hat{e}_{\rm B,K}\hat{e}_{\rm B,K'}$, which do not conserve the total number of particles in the top or bottom layers, explicitly breaking the U(1) layer symmetry.
To illustrate this, we write for these processes a similar reduced expression as in Eq.~\eqref{eqn:H_l_red} (as in Sec.~\ref{sec_SI_main_theory}, we disregard the electron spin):
\begin{equation}
    \frac{V_a}{{\cal A}}\sum_{\bm k ,\bm k'} \Big\{   \hat{\Delta}_x(\bm k) \hat{\Delta}_x(\bm k') + \hat{\Delta}_y(\bm k) \hat{\Delta}_y(\bm k') + \hat{\Delta}_z(\bm k) \hat{\Delta}_z(\bm k') - \hat{\Delta}_0(\bm k) \hat{\Delta}_0(\bm k') + \text{h.c.} \Big\}.\label{eqn:H_electron_flip_red}
\end{equation}
Clearly, these anomalous terms in Eq.~\eqref{eqn:H_electron_flip_red} break the U(1) layer symmetry.

\begin{figure}[t!]
\centering
\includegraphics[width=180mm]{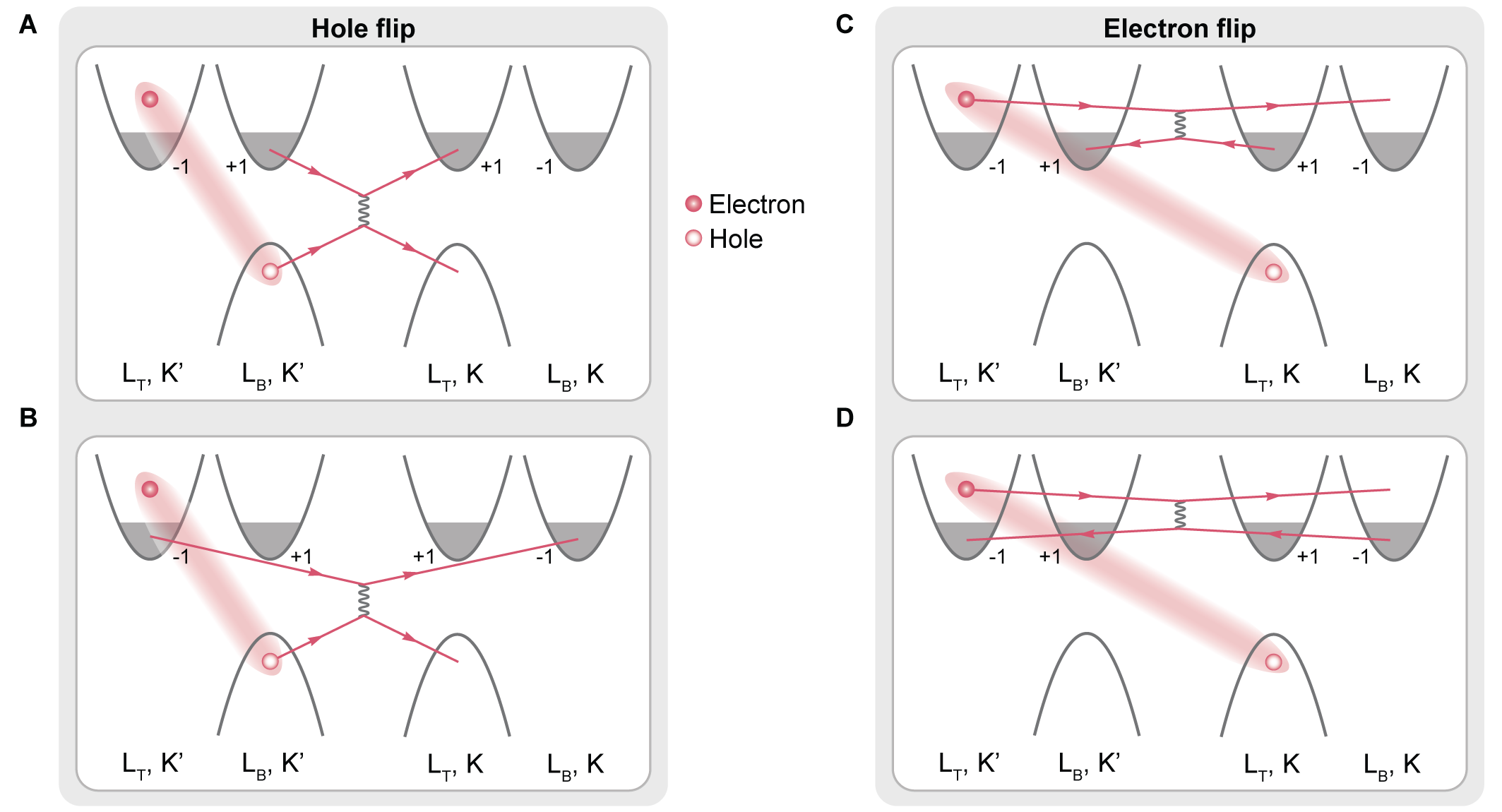} 
\caption{The hole flip (left) occurs through a scattering process, where the hole of the $X_{\rm T,K'}$-state transitions from the bottom $K'$-valley to the top $K$-valley, accompanied by one of the two depicted Fermi sea electron scatterings. Note that when an electron scatters, its spin and AQN are both preserved. The electron flip (right) occurs similarly: the electron of the intermediate $Y_{\rm T,K'}$-state transitions from the top $K'$-valley to the bottom $K$-valley, accompanied by one of the two conduction-band electron scatterings.
}
\label{fig::agnostic_flip_flop}
\end{figure}

Apart from the processes in Fig.~\ref{fig::agnostic_flip_flop}, anomalous terms generally appear in the Hamiltonian through the density-density Coulomb interactions and have the structure $\hat{e}^\dagger_{\rm T,\alpha}\hat{e}^\dagger_{\rm T,\beta}\hat{e}_{\rm B,\gamma}\hat{e}_{\rm B,\delta}$. 
These terms, involving electron layer switching (with electron density operators of the form $\hat{e}^\dagger_{\rm T,\alpha}\hat{e}_{\rm B,\beta}$), are expected to be suppressed due to wave function overlap considerations.
Assuming the AQN is a good quantum number -- supported by the small Fermi momentum relative to the lattice scale momentum, and experimentally by the absence of electron tunneling~\cite{pisoni2019absence} -- the ${\cal C}_3$-symmetry imposes that the anomalous terms take the form $\hat{e}^\dagger_{\rm T,K}\hat{e}^\dagger_{\rm T,K}\hat{e}_{\rm B,K'}\hat{e}_{\rm B,K'}$, $\hat{e}^\dagger_{\rm T,K'}\hat{e}^\dagger_{\rm T,K'}\hat{e}_{\rm B,K}\hat{e}_{\rm B,K}$, or $\hat{e}^\dagger_{\rm T,K}\hat{e}^\dagger_{\rm T,K'}\hat{e}_{\rm B,K'}\hat{e}_{\rm B,K}$.
For the first two types of terms, the involved density operators have intervalley character such as $\hat{e}^\dagger_{\rm T,K}\hat{e}_{\rm B,K'}$, so that the involved Coulomb processes carry a large momentum of the order $\bm K - \bm K'$.
Therefore, we expect that the corresponding anomalous terms are weak compared to the primary Coulomb interactions in Eq.~\eqref{eqn:H_l}. 
Indeed, a similar estimate as in Sec.~\ref{sec_SI_main_theory} for
$n = 2\times 10^{12}\,$cm$^{-2}$ gives a value $\sim V_a n/4 \simeq 1\,$meV, which is about two orders of magnitude weaker than in Sec.~\ref{subsec:estimates} if we substitute $V_a = 2\pi e^2/\varepsilon|\bm K - \bm K'|$, i.e., if we neglect the wave-function overlaps that might further suppress this estimate. 
For the third direct processes of the form $\hat{e}^\dagger_{\rm T,K}\hat{e}^\dagger_{\rm T,K'}\hat{e}_{\rm B,K'}\hat{e}_{\rm B,K}$, these interactions do not necessarily occur at large momenta near $\bm K-\bm K'$. 
However, the corresponding density operators at a finite momentum $\bm q$ have contributions from electronic states with different AQNs, such as $\hat{e}_{\rm T,K}^\dagger(\bm k + \bm q) \hat{e}_{\rm B,K}(\bm k)$. 
Due to this AQN mismatch for electrons in the same valley but opposite layers, these contributions should vanish for $q \to 0$ (otherwise, electron tunneling would be allowed, in disagreement with the experiment of Ref.~\cite{pisoni2019absence}). Hence, the anomalous terms from direct interactions are also expected to be suppressed in the low-density regime.

We finally note that the anomalous terms break the U(1) symmetry down to $\mathds{Z}_2$, i.e., the Hamiltonian remains symmetric under $\hat{\Delta}_a \to -\hat{\Delta}_a$.
Therefore, a local expectation value $\langle\hat{\Delta}_a\rangle\neq 0$ breaks this $\mathds{Z}_2$ symmetry. 
This implies that even if there are anomalous terms, which are not suppressed, we get order parameter fluctuations that involve not only the magnitude of $\langle\hat{\Delta}_a\rangle\neq 0$ but also its sign. 
As in Sec.~\ref{sec_SI_main_theory}, we, thus, expect such fluctuations to manifest in the stochastic anti-crossing.

\subsection{Second-order perturbation theory}

We write the exciton wave function as: 
\begin{align}
    \hat{X}^\dagger_{\rm T, K'} (\bm k)= \frac{1}{\sqrt{\cal A}} \sum_{\bm q} \Psi_X (\bm q) \hat{e}^\dagger_{\rm T}(\bm K' + \bm q + \bm k/2)\hat{h}^\dagger_{\rm B}(-\bm K' - \bm q + \bm k/2),\label{eqn:exc_op}
\end{align}
where $\bm k$ ($\bm q$) is the $X_{\rm T, K'}$-exciton center-of-mass momentum (momentum of the relative motion). In TMDs, the binding energy of this state is in the hundreds of meV range, and the Bohr radius $a_X$ is on the order of a few nanometers. Consequently, the momenta of both the electron and hole of this exciton are large, much larger than the Fermi momentum $k_F$. From the perspective of the $X_{\rm T,K'}$-state, the hole-flip process scatters the $X_{\rm T,K'}$-exciton into an electron-hole pair, with the hole now belonging to the top $K$-valley. Both particles are expected to have large momenta on the order of $a^{-1}_X$. We remark that the form in Eq.~\eqref{eqn:exc_op} is only approximate, as i) the $X_{\rm T,K'}$-state is expected to have appreciable spectral weight also in the $K$-valley and ii) the exciton linewidth is rather large $\gamma \simeq 10\,$meV in MoS$_2$-homobilayers~\cite{helmrich2021phonon}.

From the energy perspective, it seems most intuitive to understand the hole-flip Hamiltonian as if it couples the $\hat{X}_{\rm T,K'}$-exciton to the lowest-energy momentum-indirect exciton $\hat{Y}_{\rm T,K'}$ at the cost of perturbing the two involved Fermi seas. This excitonic $\hat{Y}_{\rm T,K'}$-state should have properties similar to the $A$-excitons, except it is optically dark.
More generally, we will consider not only the lowest-energy intralayer state but the entire Rydberg series, encompassing both bound and scattering states and fully covering the phase space of the involved electron-hole pair. The Rydberg states are expressed as:
\begin{align}
    \hat{Y}^{\dagger}_{\rm T, K',\nu} (\bm k)= \frac{1}{\sqrt{\cal A}} \sum_{\bm q} \Psi_\nu (\bm q) \hat{e}^\dagger_{\rm T}(\bm K' + \bm q + \bm k/2)\hat{h}^\dagger_{\rm T}(-\bm K - \bm q + \bm k/2).
\end{align}
Here, the index  $\nu$ runs over the entire Rydberg series, while $\bm k$ is the center-of-mass momentum -- the Rydberg series of the two-dimensional hydrogen atom is known analytically~\cite{chao1991analytical,yang1991analytic,parfitt2002two,efimkin2021electron}.

The hole-flip matrix elements can then be written as:
\begin{align}
    \hat{\cal U}(\bm p,\nu;\bm p')  \equiv \bra{Y_{\rm T,K',\nu}(\bm p)}\hat{H}_{\rm int} & \ket{X_{\rm T,K'}(\bm p')}  = -\frac{V_a}{\cal A} {\cal J}_{\nu}(\bm p,\bm p') \sum_{\bm q} \hat{\cal F}^{(h)}_{\bm q + \bm p',\bm q + \bm p}, 
    \label{eqn:hole_flip_U_a}
\end{align}
where
\begin{align}
    {\cal J}_{\nu}(\bm p,\bm p') = \frac{1}{\cal A}\sum_{\bm k} \Psi^*_\nu (\bm p' - \bm k - \bm p/2) \Psi_X(\bm p'/2 - \bm k).
\end{align}
Similarly, the electron-flip matrix elements are given by:
\begin{align}
    \hat{\cal V}(\bm p'';\bm p',\nu) \equiv \bra{X_{\rm B,K}(\bm p'')}\hat{H}_{\rm int} & \ket{Y_{\rm T,K',\nu}(\bm p')} = \frac{V_a}{\cal A} \tilde{\cal J}_{\nu}(\bm p'',\bm p') \sum_{\bm k} \hat{\cal F}^{(e)}_{\bm k - \bm p'',\bm k - \bm p'} ,  \label{eqn:V_el_flip_a}
\end{align}
where
\begin{align}
    \tilde{\cal J}_{\nu}(\bm p'',\bm p') = \frac{1}{\cal A}\sum_{\bm k} \Psi^*_X (\bm k - \bm p' + \bm p''/2) \Psi_\nu( \bm k - \bm p'/2).
\end{align}
Within second-order perturbation theory, Eqs.~\eqref{eqn:hole_flip_U_a} and~\eqref{eqn:V_el_flip_a} result in the following effective interlayer exciton hybridization of the bright excitons with $\bm p' = \bm p'' = \bm 0$:
\begin{align}
    \hat{\cal W}_{\rm inter} \approx -\frac{V_a^2}{{\cal A}^2}\sum_{\bm p,\nu} \sum_{\bm k,\bm q} \frac{\tilde{\cal J}_{\nu}(\bm 0,\bm p ) {\cal J}_{\nu}(\bm p,\bm 0) }{E_X(\bm 0) - E_\nu(\bm p) + i\gamma_\nu}   \hat{\cal F}^{(e)}_{\bm q,\bm q - \bm p}\hat{\cal F}^{(h)}_{\bm k,\bm k + \bm p}, \label{eqn:W_gen}
\end{align}
where $E_\nu(\bm p) = E_{\nu} + p^2/(2M_Y)$, $E_X(\bm p') = E_X + p'^2/(2M_X)$, and $\gamma_\nu(n,T)$ represents the decay rate of the intermediate state. We neglected the energy correction to the denominator coming from perturbing the involved Fermi seas -- such correction is expected to be small as it is set by the Fermi energy, which is typically much smaller than the energy detuning $|\delta_\nu|$, $\delta_\nu = E_\nu - E_X$. For a similar reason, the entire momentum dependence of the denominator can be disregarded. 
We note that the electron flip can occur first, followed by the hole flip, and the corresponding expression can be computed using the same analysis as outlined above.
It is instructive to rewrite Eq.~\eqref{eqn:W_gen} in the reduced form as in Eq.~\eqref{eqn:W_intra_OP} using the order parameter operators~\eqref{eqn:OP_full} (again, we disregard the electron spin):
\begin{align}
    \hat{\cal W}_{\rm inter}  \to & -V_a^2 \int\frac{d^2 \bm k}{(2\pi)^2}\int\frac{d^2 \bm k'}{(2\pi)^2} \sum_\nu \frac{\tilde{\cal J}_{\nu}(\bm 0,\bm 0 ) {\cal J}_{\nu}(\bm 0,\bm 0) }{E_X - E_\nu + i\gamma_\nu}
    \Big\{
    2 [\hat{\Delta}^\dagger_x(\bm k) \hat{\Delta}_x(\bm k') +\hat{\Delta}^\dagger_y(\bm k) \hat{\Delta}_y(\bm k')]  \label{eqn:W_inter_red}\\
    &\qquad\qquad\qquad\qquad\qquad\qquad\qquad\qquad\qquad
    + [\hat{\Delta}^\dagger_x(\bm k) \hat{\Delta}^\dagger_x(\bm k') +\hat{\Delta}^\dagger_y(\bm k) \hat{\Delta}^\dagger_y(\bm k')] + [\hat{\Delta}_x(\bm k) \hat{\Delta}_x(\bm k') +\hat{\Delta}_y(\bm k) \hat{\Delta}_y(\bm k')] 
    \Big\} \notag\\
    & 
    -V_a^2 \int\frac{d^2 \bm k}{(2\pi)^2}\int\frac{d^2 \bm k'}{(2\pi)^2} \sum_\nu \frac{\tilde{\cal J}_{\nu}(\bm 0,\bm k - \bm k' ) {\cal J}_{\nu}(\bm k - \bm k',\bm 0) }{E_X - E_\nu + i\gamma_\nu}
    \Big\{  -[\hat{\Delta}_0(\bm k) - \hat{\Delta}_z(\bm k)][\hat{\Delta}_0(\bm k') + \hat{\Delta}_z(\bm k')] \notag\\
    &\qquad\qquad
    - [\hat{\Delta}^\dagger_0(\bm k) - \hat{\Delta}^\dagger_z(\bm k)][\hat{\Delta}^\dagger_0(\bm k') + \hat{\Delta}^\dagger_z(\bm k')] + \hat{n}_{\rm B,K'}(\bm k)(1 - \hat{n}_{\rm T,K}(\bm k')) + \hat{n}_{\rm T,K'}(\bm k)(1 - \hat{n}_{\rm B,K}(\bm k')) \Big\} \notag,
\end{align}
where $\hat{n}_{\rm T,K}(\bm k) \equiv \hat{e}^\dagger_{\rm T,K}(\bm k)\hat{e}_{\rm T,K}(\bm k)$, etc.

The expression in Eq.~\eqref{eqn:W_inter_red} suggests that the expectation value $\langle\hat{\cal W}_{\rm inter}\rangle$ can be nonzero in the regime when the magnitude of the order parameter has developed but the phase is fluctuating.
This arises because the coupling in Eq.~\eqref{eqn:W_inter_red} contains terms determined solely by the magnitude of the order parameter, in contrast to the intravalley interlayer exciton hybridization in Eq.~\eqref{eqn:W_intra_OP}, which depends on the relative phase between the hole tunneling and the spatially fluctuating interlayer coherence order parameter. 
Additionally, the last two terms in Eq.~\eqref{eqn:W_inter_red} indicate that finite hybridization can occur even in a Fermi liquid state.
We note, however, that the coupling in Eq.~\eqref{eqn:W_inter_red} is expected to be small due to the factor $V_a^2$, consistent with the experimental observations in Fig.~2E,F of the main text.

\bibliography{TMD_lib}